# New scaling paradigm for dynamics in glass-forming systems

Aleksandra Drozd-Rzoska, Sylwester J. Rzoska, , Szymon Starzonek[*]

Institute of High Pressure Physics, Polish Academy of Sciences,

ul. Sokołowska 29/37, 01-142 Warsaw, Poland

[*]Corresponding author

**Highlights**

- temperature-related previtreous dynamics

- pressure related previtreous dynamics

- anomalous pressure related previtreous dynamics

- linearized distortions-sensitive analysis

- apparent activation energy index based analysis

**Abstract**

The lack of ultimate scaling relations for previtreous changes of the primary relaxation time or viscosity in glass-forming systems constitutes the grand fundamental challenge, also hindering the development of relevant material engineering applications. The report links the problem to the location of the previtreous domain remote from a hypothetical singularity. As the solution, the linearized, distortions-sensitive analysis is proposed. It is developed for scaling-relations linked to basic glass transition models: free volume, entropic, critical-like, avoided criticality, and kinetically constrained approaches. For all model scaling relations their alternative formulations based on fragility, the semi-universal metric of dynamics, are presented. The distortions-sensitive analysis is supplemented by the alternative approach based on the activation energy index showing its relative changes on cooling in the previtreous region. The search for the coherent description of the previtreous dynamics in the homologous series of polyols, from glycerol to sorbitol, is used to present in practice and validate the application of the distortions sensitive analysis. Only two scaling equation, MYEGA and the recent 'activation and critical' (AC), passed such exam. They also revealed a limited reliability of the Stickel operator' analysis for detecting the dynamic crossover. The report constitutes the unique tool-guide for applications, and a checkpoint analysis of the glass transition models. The discussion focuses on the temperature path on cooling but the extension for the still hardly discussed pressure path is also discussed.

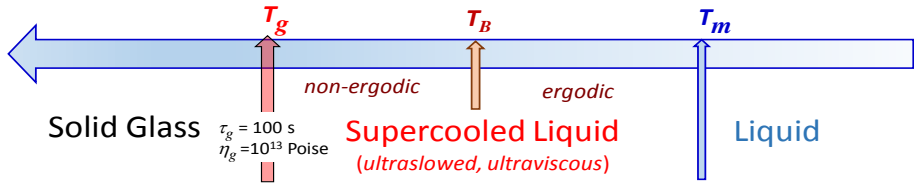

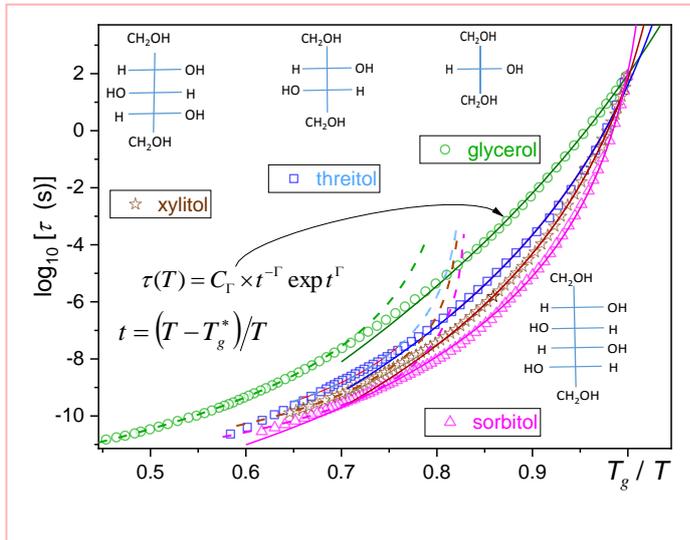

Basic polyols can vitrify at any cooling rate, passing melting temperature $T_m$ without a hallmark. It is associated with super-Arrhenius (SA), extremely fast, rise of relaxation time $\tau(T)$ or viscosity $\eta(T)$ starting even 150 K above the glass temperature $T_g$.

Note the emerging structural uniaxial symmetry in the tested homologous series.

$$\tau(T) = C_\Gamma \times t^{-\Gamma} \exp t^\Gamma$$
$$t = (T - T_g^*)/T$$

Searching for the optimal description of the previtreous behavior is amongst

**Grand Challenges of 21st Century Science**.

The plot shows normalized previtreous changes in tested polyols, portrayed by *activation-type and critical-like* (AC) relation, introduced recently.

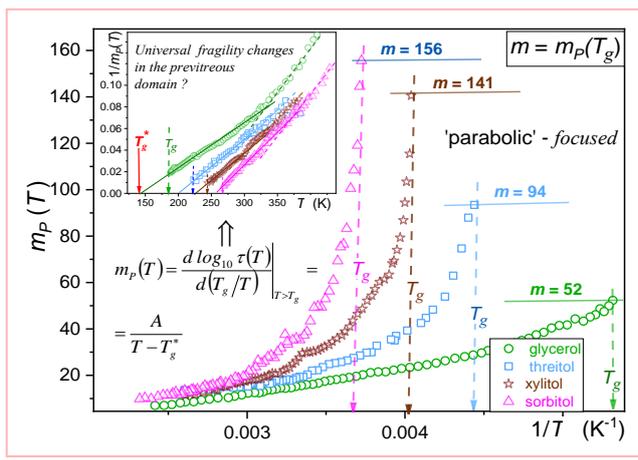

The apparent fragility $m_P(T)$ shows the steepness of $\tau(T)$ or $\eta(T)$ previtreous changes. Its value at $T_g$ is named fragility $m$ and constitute the key metric of SA dynamics. The plot also constitutes the distortions-sensitive test for the 'parabolic' scaling relation, the experimental checkpoint of Kinetically Constraint Models (KCM): its validation indicates a linear domain.

There are no such domain!

$$m_P(T) = \left.\frac{d \log_{10} \tau(T)}{d(T_g/T)}\right|_{T>T_g} = \frac{A}{T - T_g^*}$$

**The inset shows the 'empirical universality' of the apparent fragility for tested polyols.** This was the base for deriving the unique AC scaling relation (see above). The report also presents distortions sensitive and fragility based tests for the VFT, Avramov-Milchev, MYEGA, and critical-type scaling relations. Protocols for testing in prior the ability of scaling relations for describing experimental data are given. The new protocol for testing the dynamic crossover is developed.

**Table of contents**

# 1. Introduction

Vitrification and glass transition are common in nature [1-6] and essential for variety of applications: from food [7], pharmaceuticals [8], cosmetics [9], polymers [10], modern materials, chemical processing and cryogenic implementations [11-13]. It constitutes also a grand fundamental challenge for which the long-awaited cognitive breakthrough has been expected for decades. Consequently, one may rise a question why this is not happening yet?

The answer may provide a comparison with the *Physics of Critical Phenomena*, in which the descriptions of pretransitional effects approaching the singular temperature (spinodal/pseudospinodal $T_S$, or critical $T_C$) were the essential inspiration [14-18]. Related models predicted functional forms of pretransitional effects and values of relevant parameters, such as critical exponents. The dominance of collective pretransitional fluctuations was identified as a reason for a universal behavior, independent from microscopic features in a critical system. The fluctuations-related behavior is described by power terms, linked to universal critical exponents, and limited to a single term close to $T_C$ ( $\sim T_C + 1K$) [14, 15]. Such a simple description may extend even to a few tens of Kelvins for the mean-field type behavior [17-19]. It is worth to recall the case of weakly discontinuous phase transitions, where the singularity is hidden below a discontinuous phase transition temperature ($T_m$), but strong pretransitional anomalies still occurs [16-22]. One may evoke the pseudospinodal behavior in near-critical liquids or the isotropic – mesophase transitions in liquid crystalline (LC) [17-20] and plastic crystalline (PC) systems [21, 22]. They are characterized by discontinuity metric $\Delta T^* = T_m - T_S$: ranging from $\Delta T^* = 1 - 2\ K$ for the isotropic liquid – nematic transition [17-19] to even $\Delta T^* \sim 30K$ for transitions in highly ordered mesophases such a smectic E (SmE in LCs) [20] or orientationally disordered crystals (ODIC in PCs) [21, 22]. Notably, that the weak discontinuity $\Delta T^* = 1 - 2\ K$, can yield significant uncertainty in describing pretransitional effects if only the nonlinear fitting is required [16].

The glass transition is associated with pretransitional/previtreous effects starting 100 K or more above the glass temperature $T_g$, and manifesting for such a dynamic properties like viscosity $\eta(T)$ or the primary (*alpha*, structural) relaxation time $\tau(T)$. Surprisingly, after almost a century of studies, the portrayal of this phenomenon remains puzzling [23-27]. The experimental evidence indicates that previtreous effects seem to be well portrayed by a single function for a range $T_g < T < T_g + 80\,K$. Here, particular attention should be paid to the fact that glass transition is associated with discontinuity $\Delta T_g^* = T_g - T_0 = 20\,K - 50\,K$, or even more, where $T_0$ is for the extrapolated singular temperature [24-27].

This report focuses on the methodology, which may essentially reduce a parasitic bias associated with the huge discontinuity $\Delta T_g^*$ in portraying the previtreous effects.

Recalling basics of previtreous dynamics, a consensus exists for its general heuristic description as the Super-Arrhenius (SA) phenomenon, namely [25-27]:

$$\tau(T) = \tau_\infty exp\left(\frac{E_a(T)}{RT}\right) \quad , \quad \eta(T) = \eta_\infty exp\left(\frac{E_a(T)}{RT}\right) \tag{1}$$

Where $T > T_g$; $R$ stands denotes the gas constant. The essential Arrhenius dependence is retrieved if $E_a(T) = E_a = const$, in the given temperature domain.

The SA concept supported the normalized plot $log_{10}\tau(T)$ or $log_{10}\eta(T)$ vs. $T_g/T$ for common presentation of dynamics in various glass-forming liquids, introduced by Angell et al. [28, 29] and originally developed for polymers and low molecular weight liquids. Essentially, for its success was an empirical unified assumption that $\tau(T_g) = 100s$ or $\eta(T_g) = 10^{13} Poise$. Angell et al. [28, 29] also introduced an empirical metric for the universal categorization of the SA dynamics in microscopically different systems, called fragility (*m*):

$$m = m_P(T \to T_g) = dlog_{10}\tau(T \to T_g)/d(T_g/T) \quad , \quad m = dlog_{10}\eta(T \to T_g)/d(T_g/T) \tag{2}$$

In such a representation, Arrhenius dynamics is a terminal reference, manifested by a straight line between $(T_g/T = 1,\ log_{10}\tau(T_g) = 2)$ and $(T_g/T = 0,\ log_{10}\tau_\infty = -14)$. Originally, the value $\tau_\infty = 10^{-14}s$ was assumed as an estimation of a common value for the pre-exponential factor [25-29]. It yields $m = log_{10}\tau(T_g) - log_{10}\tau_\infty = 16$ for the basic Arrhenius relation. Systems with a relatively weak deviation from such a reference ($m < 40$-$50$) are referred as 'strong' glass formers. For those with $m > 50$ the class of 'fragile' glass formers with explicit SA dynamics is considered [26-29]. For the most fragile system, a limit value $m > 220$ is indicated [30, 31]. The concept of fragility has become one of the central ideas of glass transition physics [24-31]. The previtreous dynamics' changes of $\eta(T)$ and $\tau(T)$ are parallel. This report is developed mainly in terms of the latter to clarify the discussion. It is worth noting, that simple link between these magnitudes: $\tau(T,P) = (AV/k_B T)\eta(T,P)$, where $A$ is a system-dependent constant, $V$ is molecular volume, $T$ and $P$ are for temperature and pressure, respectively [32].

SA relation (Eq. 1) allows, a general cognitive analysis of the previtreous behavior, but not for the parameterization of empirical data due to the unknown form of $E_a(T)$ [24-27]. Consequently, replacement equations are required. The dominant position reached the Vogel-Fulcher-Tammann (VFT) relation, nowadays used in the form [24-30]:

$$\tau(T) = \tau_\infty exp\left(\frac{\Phi}{T-T_0}\right) = \tau_\infty exp\left(\frac{D_T T_0}{T-T_0}\right) \qquad (3)$$

where $T > T_g$, extrapolated VFT singular temperature $T_0 < T_g$ is usually located $20 - 100K$ below $T_g$; the amplitude $\Phi = D_T T_0 = const$, $D_T$ is the fragility strength coefficient describing the degree of a deviation from the basic Arrhenius pattern.

The comparison of Eqs. (1) and (2) yields the VFT formula for the apparent activation energy $E_a(T)$:

$$E_a(T) = (RD_T T_0)[(T-T_0)/T]^{-1} = (RD_T T_0)t^{-1} = Et^{-1} . \ E = const \qquad (4)$$

The hypothetical universality of the VFT relation supported applications for determining glass transition characteristics based on $\tau(T)$ or $\eta(T)$ analysis remote from $T_g$. For instance [29, 33]:

$$m = \frac{D_T T_0 T_g}{(\Delta T_g^*)^2 ln10} \quad \text{or} \quad m = \mu\left(1 + \frac{ln10}{D_T}\right) \Rightarrow D_T = \frac{\mu ln10}{m-\mu} \qquad (5)$$

where $\Delta T_g^* = T_g - T_0$, $\mu = m_{min} = log_{10}\tau(T_g) - log_{10}\tau_\infty$: assuming $\tau_\infty = 10^{-14}s$ one obtains $\mu = 16$.

For macromolecular systems, Williams-Landell-Ferry (WLF, 1955) relation is more convenient for basic experimental methodologies [34]. However, WLF and VFT equations are isomorphic and the discussion may focus on the latter [26, 27]. VFT equation is related to three adjustable parameters: pre-exponential factor $\tau_\infty$, extrapolated singular temperature $T_0$ and fragility strength $D_T$. This number is recognized as optimal for scaling relations describing the previtreous dynamics [24-29].

The extensive experimental evidence supporting the meaning of the VFT/WLF relations made them an empirical symbol of the previtreous dynamics universality [23-30]. Consequently, the derivation of VFT relation became a check point for theoretical models. However, none of the glass transition models manage to derive specific values of parameters in this equation [24-27, 35-37]. This trend also strongly supports numerous experimental results suggesting coincidence between the temperature $T_0$ determined from dynamic studies and the Kauzmann temperature $T_K$ (ideal glass) determined from the configuration entropy analysis, i.e., the thermodynamic insight [15-18, 38, 39].

However, the state-of-the-art analysis for 52 glass-formers carried out by Tanaka [40] showed that $0.8 < T_0/T_K < 2.2$, so the correlation $T_0 \approx T_K$ can be considered only selected systems. Further, precise tests of the fragility determined via Eq. (5) revealed notable discrepancies with the direct estimation of fragility index based on the 'Angell plot' [41, 42]. Decisive arguments questioning universality and fundamental significance of the VFT equation delivered the analysis basing on activation energy index (see below) [43-47]. Being inspired by these results, McKenna conducted a subtle study of the previtreous effect in polymers using the WLF relation and showed systematic rise of deviations on cooling towards $T_g$ [48]. These results led to questioning fundamental significance of VFT and WLF relations [44, 45]. After decades of research, the most crucial experimental fact in the physics of glass transition turned out to be unknown. It seemed that only new 3-parameter model-equations could overcome this cognitive deadlock [44].

The principal method of verifying new scaling equations is a visual or quantitative (residual) comparison of fitting quality for different experimental data sets [49-56]. However, such a classic analysis did not lead to a decisive prevalence of one model equation over another. This confusion may explain the non-accessibility of a domain in the vicinity of the singular temperature $T_0$ with the most characteristic changes of $\tau(T)$ or $\eta(T)$, due to the very large value of the discontinuity $\Delta T^*$. Another way to prove the adequacy of a given scaling relation a plot linking several dozen sets of experimental data for different glass-forming systems and obtaining a single scaling curve within overlapping data sets [57-63]. Such plots often serve as crucial empirical validations of theoretical model matched to the scaling relation. However, the analysis is performed by the use of three-parameters scaling relation and calculated individually for each selected system. Therefore, it is a kind of 'tautological validation', the success of which is guaranteed in advance. But even with such a inconclusive results, one may find that some equations offer a subtly better fit than others. Unfortunately for other glass-forming systems, the situation seems to be flipped. Hence, if a universal scaling equation the previtreous dynamics exist, a question rises?

This report presents the methodology for analyzing previtreous effects that may respond to above challenges. It bases on linearized differential analysis sensitive to subtle disturbances between a scaling relation and experimental data. Contrary to the common practice, experimental validations tests are focused on a portrayal within a homologous series of glass-formers, with a systematic change of molecular structure. Finally, the state-of-the-art analysis of the pressure path approaching the glass transition is discussed. It includes new development for this issue.

## 2. VFT relation and its links to basic glass transition models

In 1889 Arrhenius proposed the empirical formula for the temperature dependence of chemical reaction rates $k(T) = k_\infty exp(E_a/RT)$ [64] This relation introduced the concept of a process activation energy, easily determined from the linearized plot $lnk(1/T) = lnk_\infty + E_a(1/T)/R$. It became the base for describing many thermally induced dynamic processes in physical chemistry, including viscosity (Guzman, 1913 [65], Raman 1923 [66] and Andrade, 1933 [67]), primary relaxation time (Williams, 1964 [68, 69]) as well as diffusion or the electric conductivity changes. In the 20$^{th}$ century, the industrial revolution entered stage where the detailed description of such a behavior became important for technological implementations. However, it became clear that the behavior goes beyond the Arrhenius pattern for many systems. Vogel (1921, [70] for mineral oils (lubricants, fuels, petrochemical industry) as well as Tammann [71], and Fulcher [72], in response to the challenges of the glass industry, introduced an additional parameter to the Arrhenius relation creating the enhanced 'functional flexibility': $log_{10}\eta(T) = A + B/(T - \theta)$. For example, with the help of this relation, Fulcher was able to successfully describe changes in the viscosity of soda-lime glasses with various compositions ranging from 500 to even 1400ºC.

Universality of VFT and WLF relations in subsequent decadence caused its derivation to become the checkpoint for glass transition models [24-29, 49]. Doolittle [73], Turnbull and Cohen [35] considered the free volume concept allowing for a molecule, or a polymer segment [74] movement, reaching the output relation:

$$\tau = \tau_\infty exp\left(\frac{\gamma v^*_{free}}{v_{free}}\right) = \tau_\infty exp(\gamma\phi) \qquad (6)$$

where $v^*_{free}$ is a minimum required volume of void for reorientation process, $\gamma$ is an overlap factor that should lie between 0.5 and 1, and $v_{free}$ is specific free volume; fraction coefficient: $\phi = v^*_{free}/v_{free}$.

Eq. (6) converts into the VFT Eq. (3) assuming $\phi = A(T - T_0)$

Adam and Gibbs [36] visualized a supercooled liquid as progressively self-organized cooperatively rearranging region (CRR), which arrangement is inversely proportional to the configurational entropy $S_C$. Hence, the configurational entropy $S_C(T)$ decreases on cooling, which is coupled to the increase of particles in CRRs. The primary relaxation time is interpreted as the rate needed to rearrange the region, and its evolution is expressed by model output relation [36]:

$$\tau(T) = \tau_\infty exp\frac{z^*\Delta\mu}{k_B T} = \tau_\infty exp\frac{B}{TS_C(T)} \qquad (7)$$

where $\Delta\mu$ denotes transition state activation energy, $z^*$ is temperature-dependent number of cooperatively rearranging molecular entities determined by macroscopic configurational entropy $S_C$, in such a way that $z^*/s^*_c = N_A/S_C(T)$, in which $s^*_c$ stands for the entropy of the smallest number of rearranging molecular entities and $N_A$ is the Avogadro number.

Experimentally, configurational entropy $S_C(T)$ is estimated from the evolution of the heat capacity excess $\Delta C_P(T)$ [31, 47, 50]:

$$S_C = \int_{T_K}^{T} \frac{\Delta C_P(T)}{T} dT \qquad (8)$$

In practice, it is assumed $\Delta C_P(T) = C_P^{Superc.Liq.} - C_P^{Solid\ Glass}$, with heat capacity of glass, instead of hardly detectable for glass formers, solid crystal entropy changes.

VFT equation is retrieved if following behavior for heat capacity and configurational entropy is assumed [26, 27, 49, 75]:

$$\Delta C_P(T) = \frac{\Delta C_P}{T} \quad \Rightarrow \quad S_C(T) = S_0\left(1 - \frac{T_K}{T}\right) = S_0\left(\frac{T-T_K}{T}\right) = S_0 t \tag{9}$$

The random first-order field theory (RFOT), also known as the mosaic theory, assumes the nucleation of 'entropic droplets' between different metastable configurations, creating a patchwork of local metastable configurations in a supercooled liquid [49, 76]. It predicts the link between the primary (*α, structural*) relaxation time, static length scale $\xi(T)$ and configurational entropy $S_C(T)$ [75, 112]:

$$\tau(T) = \tau_\infty exp\left(\frac{A\xi^\psi}{T}\right) \quad \text{and} \quad \xi(T) \propto \left(\frac{1}{S_C(T)}\right)^{\frac{1}{d-\theta}} \tag{10}$$

where $d$ is spatial dimension, $\theta$ is an exponent related to an interface energy ($Y$) changes between two amorphous states. The exponent $\psi$ is related to a free energy barrier to overcome when rearranging a correlated volume of $\xi$ size. Model-values linking exponent to a specific glass former have not been computed yet. Comparing Eqs. 7 and 10, the enhanced AG model equation emerges [112]:

$$\tau(T) = \tau_\infty exp\left(\frac{A_G}{T(S_C)^\alpha}\right) \tag{11}$$

where the exponent $\alpha = \psi/(d-\theta)$.

This basic RFOT relation can be reduced to VFT dependence if the exponent $\psi = d - \theta$ and $TS_C(T) = (T - T_K)$.

Tanaka [37] considered a pretransitional behavior as a critical-like and Ising-like phenomenon and derived the relation $\tau(T) = \tau_\infty exp D_T(\xi(T)/\xi_\infty)^{d/2}$, where the correlation length $\xi(T) = \xi_\infty(T - T_K)^{\nu=2/d}$ is associated with the singularity at the extrapolated Kauzmann temperature.

As described above, basic problems of VFT parameterization inspired by development of alternative scaling relations, particularly without the finite temperature divergence. The leading position seems to gain Mauro-Yue-Ellison-Gupta-Allan (MYEGA) dependence [52], for which configurational entropy is considered within topological constraint model as $S_C(T) = f_t(T)Nk_B \ln \Omega$, where $N$ is the number of species (atoms, molecules), $k_B$ is the Boltzmann factor, and $\Omega$ is the number of degenerate configurations per floppy mode. Further, the two-state model for topological degrees of freedom, the two-state model, in which network constraints are either intact or broken, with an energy difference given by *H(T)*, which may be related to enthalpy, was applied, yielding: $f_t(T) = 3\exp(-H/k_B T)$. Substituting all these into AG Eq. (12) MYEGA relation was obtained [52]:

$$\tau(T) = \tau_\infty \exp\left[\frac{C}{T}\exp\left(\frac{K}{T}\right)\right] \tag{12}$$

where $K = A_{AG}/3Nk_B \ln \Omega$ and $C = H/K$ are constants.

It can be approximated by VFT dependence assuming validity of first-order term Taylor series expansion [78]:

$$\ln\left(\frac{\tau(T)}{\tau_\infty}\right) = \frac{C}{T}\exp\left(\frac{K}{T}\right) = \frac{C}{T\exp(-K/T)} \approx \frac{C}{T(1-K/T)} = \frac{C}{T-K} \tag{13}$$

This yields the VFT equation for $K = T_0$ and $C = D_T T_0$.

In the 1980s, more detailed studies indicated some systems problems with the state-of-the-art portrayal of experimental data by VFT relation. The first remedy for these problems was to introduce the exponent [79, 80]:

$$\tau(T) = \tau_\infty \exp\left(\frac{A}{T-T_0}\right)^w \tag{14}$$

Such a description is used within a few theoretical models predicting the well-defined value of *w*. One of the most popular, proposed by Bendler and Schlezinger [79], linked the ultraviscous/ultraslowing behavior to appearance of local and temporary mobile defects and reached the SA type relation, in the type of Eq. (4)) with $w = 3/2$.

The above resume shows the essential significance of VFT portrayal of previtreous effect for basic glass transition model. The problem emerges, when taking into account references recalled at the end of the Introduction section, questioning two essential paradigmatic assumptions: (i) correlation between (dynamic singularity) and $T_K$ (thermodynamic singularity) is doubtful, (ii) VFT equation seems to offer only an effective portrayal, limited to selected systems. Notwithstanding, usage of VFT equation in experimental, theoretical, and practical applications has been grown up permanently in the last decades, as illustrated in Figure 1.

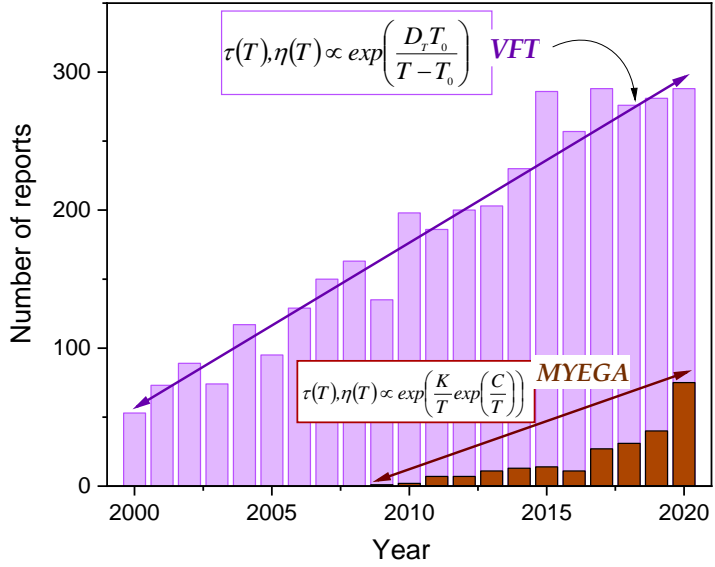

*Fig. 1. Number of reports presenting VFT relation in the last two decades. It also shows number of works using the MYEGA relation, which has been an important alternative description of the pre-vitrification dynamics since 2010. Results were obtained using Google Scholar.*

When discussing experimental validation of VFT Eq. (3), it is worth to recall also analysis proposed by Stickel, used to detect the dynamic crossover temperature $T_B$ [80]. It is realized by transformation of experimental data $\tau(T) \rightarrow \varphi_T = [d\,log_{10}\,\tau(T)/d(1/T)]^{-1/2}$, and plot $\varphi_T(T)$ vs. $1/T$, yielding two lines intersecting at dynamic crossover temperature $T_B$. The latter separates ergodic (high-temperature) and non-ergodic (low-temperature) dynamic domains in the ultraviscous region [49, 80, 81]. In both domains, another optimal evolution of dynamic properties is predicted. For the HT one, the description is very similar, for $T > T_B + (10K \div 20K)$, with an extrapolated singular temperature $T_C^{MCT} \approx T_B$ [49, 61, 82, 83]:

$$\tau(T) = \tau_0(T - T_C^{MCT}) \tag{15}$$

The description in HT domain using relation (15) is well-founded within the mode-coupling theory (MCT). Usually $T_B \approx 1.3 T_g$, which estimates range of LT domain to $80K \div 100K$. Novikov and Sokolov [84] strengthened the possible fundamental significance of $T_B$ and $\phi_T(T)$ plot, by announcing the semi-universal 'magic' time scale $\tau(T_{B\rightleftarrows}) = 10^{-7\pm 1}s$. This empirical finding was obtained by the use of $\varphi_T(T)$ plots for 29 glass-forming low-molecular-weight liquids, polymers, ionic systems, covalent systems, and plastic crystals [84]. Notwithstanding, a few striking discrepancies from 'magic' time-scale has been noted later [32, 86, 87, 88]. Casalini and Roland [89] developed above concept for pressure path on approaching glass transition via plot $(dlog_{10}\tau(P)/dP)^{-1/2}$ vs. $P$ and indicated empirical invariance $\tau(T_{B\rightleftarrows}, P_B) = 10^{-7\pm 1}s$ in the pressure-temperature plane. However, the question arises as to the Stickel analysis, closely related to VFT relation, is influenced by fundamental doubts about the latter? Another issue is that Stickel et al. analysis assumed the preference for VFT description in both dynamic domains, while the experimental evidence indicates a universal tendency to prefer MCT Eq. (15) in HT dynamic domain.

## 3. Experimental

Studies were carried out in homologous series of polyols, from glycerol to sorbitol, for which changes in molecular structure lead to the emergence of the uniaxial molecular symmetry. They belong to classic glass-forming systems, which hardly crystallize when passing the melting temperature [29, 49, 90]. Such a feature facilitates broadband dielectric spectroscopy (BDS) studies, requiring frequency scans of the electric impedance, lasting several minutes or more for $T \to T_g$ [91]. Compounds were purchased from Sigma-Aldrich, anhydrous, analytic quality grade and used without additional purification. The measurement capacitor was filled in a dry box. The gap of the capacitor $d = 0.2mm$ and the voltage of the applied electric field $U = 1V$ were performed. The Quattro automatized unit supplemented the Novocontrol impedance analyzer for temperature control was used. These enabled the five-digit permanent resolutions for parameters, and the temperature stability was better than 0.1 K.

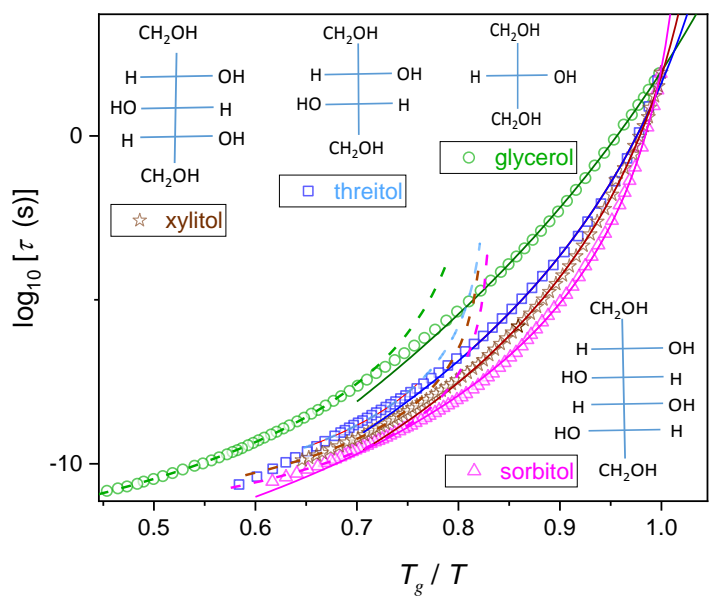

*Fig. 2 Normalized Angell plot for previtreous changes of primary relaxation time in polyols: glycerol ($T_g$ = 187.7 K), threitol ($T_g$ = 224.4 K), xylitol ($T_g$ = 247.6 K), and sorbitol ($T_g$ = 268.3 K) [25, 91]. Schematic structures of compounds are shown. Solid (low-temperature, non-ergodic dynamic domain) and dashed curves (high temperature, ergodic domain) are related to activation-critical relation Eq. (31), parameters are given in Table I.*

The research was focused on the primary (*alpha, structural*) relaxation time, determined using derivative analysis of primary loss curve to find location where $d\, log_{10}\, \varepsilon''(f = f_{peak})/d\, log_{19} f = 0$ what yields relaxation time $\tau = 1/2\pi f_{peak}$. Such an approach minimizes the fitting uncertainty, which may be significant for the most popular way of multi-parameter fitting of $\varepsilon''(f)$ loss curve using the Havriliak-Negami or related functions [16, 77]. Obtained evolutions of primary relaxation time in tested compounds are shown in Figure 2, using normalized Angell plot presentation. For such a plot, Arrhenius behavior is referred to as the linear one. Increasing curvature shows rising degree of SA behavior, characterized by fragility index *m*. It changes from $m \approx 52$ for glycerol to $m \approx 158$ for sorbitol, in agreement with earlier estimations [25, 91]. This rise correlates with emergence of uniaxial structure or increasing importance of hydrogen bonding [25, 90, 91, 92, 93]. Generally, the previtreous dynamics of polyols are discussed concerning essential role of hydrogen bonding. Emergence of local preferably uniaxial structures increases number of neighboring molecules, which yields more possibilities for hydrogen bonding and densifying -OH groups locally. A similar impact of uniaxial form of molecules is well known in isotropic liquid phase of rod-like liquid crystals. Consequently, there is no contradiction between increasing role of hydrogen bonds and less frequently discussed uniaxiality of molecular structure in discussed polyols. Instead, a synergy between mentioned factors may be expected.

## 4. Linearized, derivative-based analysis of the previtreous dynamics

Previtreous changes of dynamic properties, such as viscosity or primary relaxation time, start even more than 100 K above the glass temperature [25]. Their analysis via scaling relations is possible only remote from singular temperatures, such $T_0$ in VFT Eq. (3). One may introduce discontinuity metric $\Delta T_g^* = T_g - T_0$, which can even exceed 50 K [25, 40, 45]. It estimates domain in the vicinity of singular temperature, which is non-accessible for analysis of $\tau(T)$ or $\eta(T)$ evolutions. Unfortunately, it is also a region with the most characteristic changes of these properties. The experimentally accessible region $T > T_g = T_0 + \Delta T_g^*$ can be considered as the specific 'long tail' of previtreous changes. Consequently, validations test of scaling-relations based only on the visual or residual analysis cannot be decisive.

This cognitive impasse may be overcome by linearized distortions-sensitive analysis presented in this section. It bases on the distortions-sensitive transformation of experimental data, yielding linear behavior in the domain, where given scaling relation may be applied. The significant problem of scaling relations derived from glass transition models is that the latter does not yield specific values of parameters linked to given experimental system [24-27, 35, 36, 37, 49, 52, 59, 76, 77, 79 ]. To reduce this problem, model scaling relations are also presented in terms of fragility, the surrogate empirical parameter characterizing potential universality of dynamics in glass formers. Fragility constitutes a terminal of apparent fragility $m_P(T \leq T_g)$, also known as the steepness [49], which is directly linked to apparent activation enthalpy $H_a(T)$:

$$\tau(T) \rightarrow H_a'(T) = \frac{H_a(T)}{R} = \frac{d \ln \tau(T)}{d(1/T)} = \left(\frac{T_g}{\log_{10} e}\right) \frac{d \log_{10} \tau(T)}{d(T_g/T)} = (T_g \ln 10) m_P(T) \qquad (17)$$

The discussion of basic model-relations starts from MYEGA Eq. (12), often indicated as a possible successor of VFT dependence grand success. It is finished by issues related to the basic SA-type VFT and critical-like (CL) portrayals and the very recent relation linking both these approaches. As described above, basic problems of VFT parameterization inspired development of alternative scaling relations, particularly without finite temperature divergence. Leading position seems to gain Mauro-Yue-Ellison-Gupta-Allan (MYEGA) dependence [52]. Eq. (12) can also be expressed in fragility-related characteristics, namely $K = \mu T_g = const$ and $C = [(\mu - 1)T_g/T] - 1 = [(\mu - 1)T_g - T]/T$ [52].

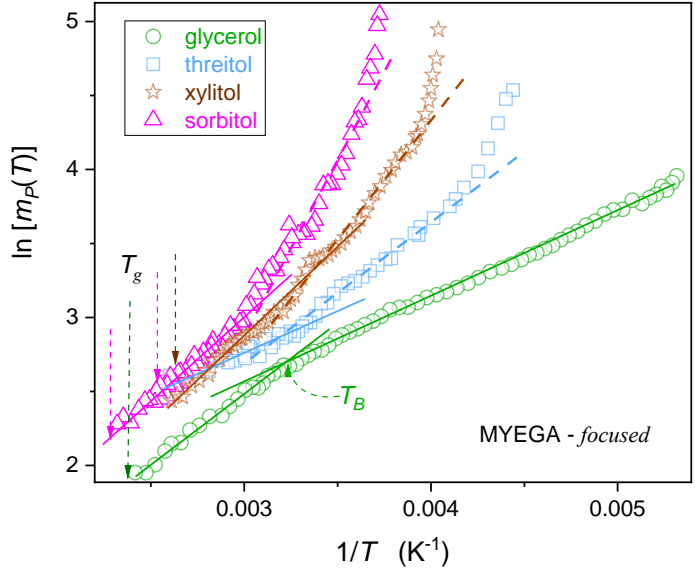

*Fig. 3 Distortion-sensitive tests for preferable portrayal $\tau(T)$ previtreous changes in polyols via MYEGA relation [52] (Eqs. (12)): such domains should follow linear pattern. Intersections of straight lines indicate possible dynamical crossover temperatures: (i) glycerol $T_B = 312$ K, (ii) threitol $T_B = 315$K, (iii) xylitol $T_B = 300$ K, and (iv) sorbitol $T_B = 295$ K.*

Note that Eq. (12) may be alternatively derived from basic free volume assuming $f = C'T\, exp(K/T)$. Focusing on the linearized, distortions-sensitive test for MYEGA dependence, one obtains:

$$\frac{d \ln \tau(T)}{d(1/T)} = H'_a(T) = C \exp\left(\frac{K}{T}\right) + \frac{CK}{T}\exp\left(\frac{K}{T}\right) = \exp\left(\frac{K}{T}\right)\left[C + \frac{CK}{T}\right] \to \ln H'_a(T) = \frac{K}{T} + \ln\left[C\left(1 + \frac{K}{T}\right)\right] \approx$$

$$\frac{2K}{T} + \ln C = A\frac{1}{T} + B \tag{18}$$

Linear domain in plot $H_a^{'}(T)/exp(1/T)$ vs. $1/T$ indicates region of applicability of MYEGA description. The linear regression yields parameters: $K = B/A$, $C = A/exp\,K$.

Figure 3 shows tests for distortion-sensitive analysis via Eq. (18) of MYEGA Eq. (12). Evident linear domains appear for the low-temperature and high-temperature dynamical domains. It indicates MYEGA Eq. (18) possibilities to portray previtreous dynamics and a new way to test the dynamic crossover phenomenon.

Over three decades ago, Avramov and Milchev derived another scaling relation for the previtreous dynamics avoiding the finite temperature singularity [53]:

$$\tau(T) = \tau_\infty exp\left(\frac{A_{AM}}{T^D}\right) \qquad (19)$$

Applying fragility concept, one may express this dependence in the 'universal' form $D = m/\mu$. Notably, link to earlier Bässler equation for which $D = 2$ [94]. Comparing AM Eq. (19) with Eqs. (6) and (7), one obtains $f = C/T^D$ for free volume fraction and $S_C \propto T^{D-1}$ for configurational entropy, in disagreement with experimental evidence [49, 75]. Considering linearized, derivative-based test for AM Eq. (19), one obtains [87, 95]:

$$log_{10}\left[\frac{d(ln\,\tau)}{d(1/T)}\right] = log_{10}H_a^{'} = log_{10}(CD) + (1-D)\,log_{10}T = A + B\,log_{10}T \qquad (20)$$

For plot $log_{10}H_a'(T)$ vs. $log_{10}T$, linear behavior indicates region of applicability of AM relation and linear regression yields optimal values of parameters: $D = 1 - B$ and $C = 10^A/(1-B)$.

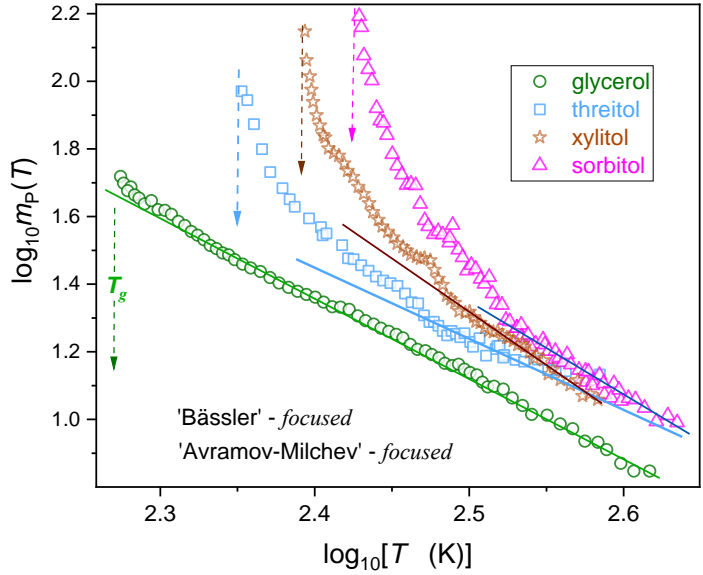

*Fig. 4. Distortion-sensitive plot focused on searching domains, preferably described by the AM [53] Eq. (19): straight lines indicate them.*

Figure 4 presents linearized, distortions-sensitive test of Avramov-Milchev [53] and Bässler [94] equations. The latter is considered as one of possible output relations for Kinetic Constraint Models (KCM) [96]. For the broader range of temperatures, the preference for AM Eq. (19) portrayal appears only for glycerol, but with the parameter $D \neq 2$.

In the last decade, notable efforts were devoted to dynamic facilitation theories (DFT), particularly within KCM models frames [59, 76, 77, 96-98]. They considered vitrification as a purely kinetic phenomenon, for which movements of molecules in previtreous, supercooled regions are associated with excitations that appear/disappear in the adjacent areas and these facilitated dynamics develop in a hierarchical and correlated fashion in a specific direction. This picture emerges when cooling below inset temperature, associated with the Arrhenius–non-Arrhenius crossover. Chandler, Garrahan, and Elmatad (CGE) [59, 77] derived the basic experimental checkpoint relation for KCM approaches, obeying between the high-temperature Arrhenius – non-Arrhenius onset temperature $T_o$ and the glass temperature $T_g$ [59]:

$$log_{10}\left(\frac{\tau(T)}{\tau_o}\right) = \left(\frac{J}{T_o}\right)^2 \left(\frac{T_o}{T} - 1\right)^2 \quad \Rightarrow \quad \tau(T) = \tau_o exp\left[G\left(\frac{1}{T} - \frac{1}{T_0}\right)^2\right] \qquad (21)$$

where $G = J^2/ln10$.

For a large enough value of $T_o$, it may be approximated the Bässler equation which appears in the East model within simplified DFT approach [96-98]. For the linear derivative-based test of CGE Eq. (21) the following relation can be derived:

$$m_P(T) = \frac{2J}{T_g}\left(\frac{1}{T} - \frac{1}{T_o}\right) = A\frac{1}{T} + B \qquad (22)$$

For the plot $m_P(T)$ or $H'_a(T)$ vs. $1/T$ linear behavior, validates application of CGE Eq. (21) for the given glass former and temperature domain occurs. The subsequent linear regression yields optimal values of parameters $J = A/2T_g$, $T_o = A/B$. One can also present Eq. (21) in terms of the 'universal' metric, fragility, substituting $J^2 = (m/2)[T_g/(T_o/T_g - 1)]$, as results from Eq. (22) for $T = T_g$. Eq. (21) is often recalled as the 'parabolic relation' for describing previtreous dynamics, due to the Bässler-type approximation: $log_{10}\tau(T) \propto 1/T^2$ [96-98]. Iverlapping of $\tau(T)$ experimental data for 68 glass-forming systems in ref. [59], is recalled as a crucial argument supporting universal meaning of CGE Eq.(21) and the experimental validation of DFT/KCM models. In the opinion of the authors, this experimental result (shown in ref. [59]) has tautological features and cannot be considered as a conclusive validation of CGE Eq. (21), since the basic plot is scaled: $log_{10}(\tau/\tau_o)$ vs. $(J/T_o)^2(T_o/T - 1)^2$, i.e., using all adjustable parameters included in Eq. (21), individually for each selected glass-formers. Similarly, 3-parameters based scaling plots showing overlapping of experimental data are known for other scaling relations, for instance: $(\tau(T)/\tau_0)^{-1/\phi}$ vs. $T/T_C$ for CL Eq. (27) [61], $log_{10}\tau$ vs. $A_{VFT}/(T - T_0)$ for VFT Eq. (3) [25, 60, 99, 100], $log_{10}\tau$ vs. $A_{AM}/T^D$ for AM Eq. (19) [53]. One may also propose a plot for MYEGA Eq. (12): $T\,log_{10}\tau$ vs $C'\,ln(K/T)$. In each case, such a scaling plots *a priori* lead to overlapping all used experimental data, if only given relation may effectively portray results data within the limits of experimental errors. Hence, such multiple scaled plots cannot be considered as a validation for a given model relation.

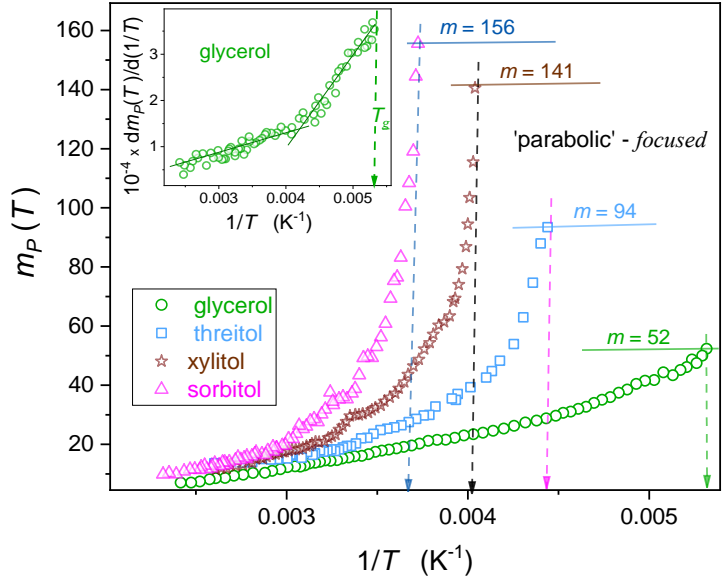

*Fig. 5. Temperature evolution of apparent fragility (steepness index), in polyols, using scale correlated to Angell plot and Eq. (21). The latter is focused on testing the preference for CGE description [57]: such domains should follow linear pattern. The inset presents derivative data from the central part of the plot, indicating that evolution $m_P(1/T)$ is not linear for glycerol.*

Figure 5 shows that for a member of tested series of polyols, changes of $m_P(T)$ are strongly non-linear, including glycerol, for which additional test is shown in the inset. Following Eq. (21) the results presented in Fig. 5 show fundamental inadequacy of 'parabolic' CGE Eq. (21) [57, 76, 77] for portraying previtreous dynamics in glycerol, threitol, xylitol, and sorbitol.

As mentioned above, the VFT Eq. (3) may be considered as dominant relation for portraying previtreous dynamics. It can be alternatively presented using the fragility:

$$log_{10} \tau(T) = log_{10} \tau_\infty + \frac{\varsigma^2}{m(T/T_g - 1) + \mu} \tag{23}$$

where $\varsigma = m \, log_{10} \tau(T_g) \, log_{10} \tau_{\infty \, min.}$

The comparison of Eqs. (3) and (23) yield the link: $D_T = T_g m_{min}^2 / m$ and $T_0 = T_g \left(1 - m_{min.}/m(T/T_g - 1)\right)$. For VFT relation, one may derive following linearized equation [87]:

$$\left[\frac{d \ln \tau(T)}{d(1/T)}\right]^{-1/2} = [H_a'(T)]^{-1/2} = (D_T T_0)^{-1/2} - T_0(D_T T_0)^{-1/2} \times \frac{1}{T} = -A\frac{1}{T} + B \qquad (24)$$

$$T[H_a'(T)]^{-1/2} = (D_T T_0)^{-1/2} \times T - T_0(D_T T_0)^{-1/2} = AT - B \qquad (25)$$

For plots $[H_a'(T)]^{-1/2}$, $[m_P(T)]^{-1/2}$ vs. $1/T$ or $T \times [H_a'(T)]^{-1/2}$, $T \times [m_P(T)]^{-1/2}$ vs. $T$ the linear behavior indicates domains where VFT equation can be used. Linear regression fit may yield parameters $A$ and $B$, and consequently: $D_T = 1/AB$ and $T_0 = B/A$. Notably, the link between Stickel operator $\varphi_T$ [80] analysis, originally focused on determining dynamic crossover temperature:

$$\left(\frac{d \ln \tau(T)}{d(1/T)}\right)^{-1/2} = [H_a'(T)]^{-1/2} = (T_g \ln 10)^{-1/2}[m_P(T)]^{-1/2} = \frac{1}{\sqrt{\ln 10}}\varphi_T(T) \qquad (26)$$

As mentioned above, in high-temperature dynamic domains of supercooled systems, critical-like portrayal within the MCT approach is advised. The same type of portraying was considered for the low-temperature dynamic domain, close to the glass temperature:

$$\tau(T) = \tau_0'(T - T_C)^{-\varphi} \quad , \quad \tau(T) = \tau_0 \left(\frac{T - T_C}{T_C}\right)^{-\varphi} \qquad (27)$$

$$\tau(T) = \tau_0 \left(\frac{T - T_C}{T}\right)^{-\varphi} \qquad (28)$$

where $T \geq T_g$, $T_C < T_g$. For dynamic critical phenomena [101]: $\varphi = z\nu$; $\nu$ and $z$ is exponents for correlation length $\xi$, and $z$ is a dynamic exponent.

Four decades ago, Souletie and Bertrand carried out comparative tests of such descriptions for several systems [102, 103]. Unfortunately, their results show the non-conclusive scatter of the exponent $\varphi$ and the relatively poor-fitting quality. Saltzmann and Schweitzer [104] analyzed numerically a hypothetical critical universality in polymeric glass formers and suggested $\varphi \approx 1.7$. Experimental validation of these results in low molecular weight liquids seems doubtful [59, 90]. Two decades ago, Colby [56] announced hypothetically breakthrough results, indicating universal and critical-like behavior by Eq. (27) with the universal exponent $\varphi = z \times \nu = 6 \cdot 3/2 = 9$, supported by validating evidence for 35 glass-forming systems [56]. The exponential multiplicator was advised for some molecular liquids [57]. Heuristic considerations supporting this reasoning were called as the 'dynamic scaling model' (DSM) [56, 57]. However, this result has been skeptically treated because even for the same experimental datasets as in ref. [56, 57] declared universal DSM criticality was not confirmed [55].

Authors of this report developed the linearized distortions-sensitive analysis for the critical-like Eq. (19) [87]:

$$\frac{T^2}{H'_a} = \frac{T_C}{\varphi} - \varphi^{-1}T = A - BT \tag{29}$$

Using a plot $T^2/H'_a(T)$ vs. $T$, linear behavior shows domains where Eqs. (27) and (28) can be applied; subsequent linear regression yields basic parameters: $T_C = AB$ and $\varphi = 1/B$.

In refs. [45, 86, 105-109] linearized distortion-sensitive analysis was applied for liquid crystalline glass-formers composed of rod-like molecules, showing clear prevalence for critical-like portrayal with the exponent $\varphi = 9$. Hence, behavior suggested by DSM approach appears, although not in systems indicated in refs. [56, 57]. The prevalence of critical-like portrayal was also found in plastic crystals and some low-molecular-weight liquid and polymers where local elements of uniaxial symmetry occurs [45, 86].

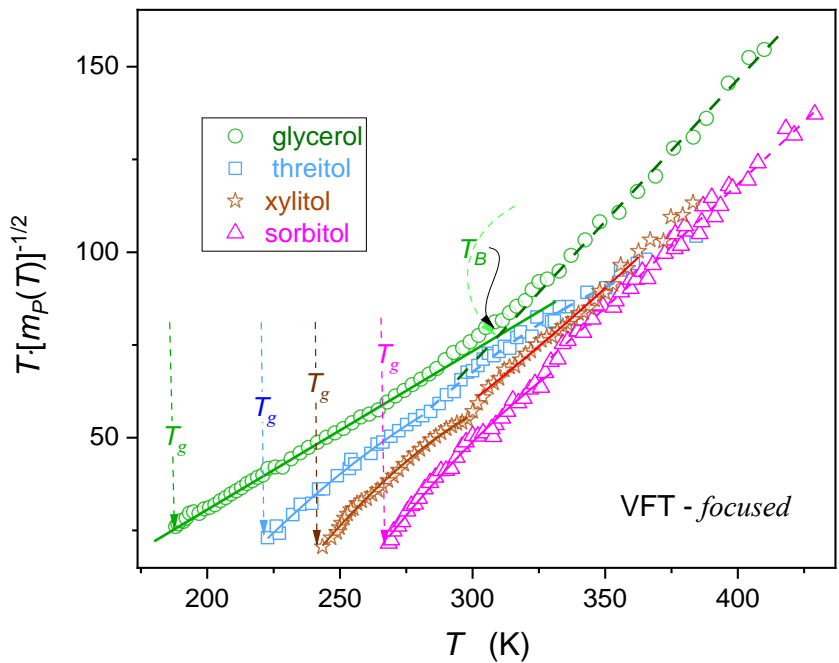

*Fig. 6. Distortion-sensitive plot (Eq. (26)) focused on searching domains, preferably described by VFT description Eq. (3): straight lines indicate them. The presentation is equivalent to the Stickel et al. [80] plot introduced for detecting the dynamic cross-over temperature $T_B$.*

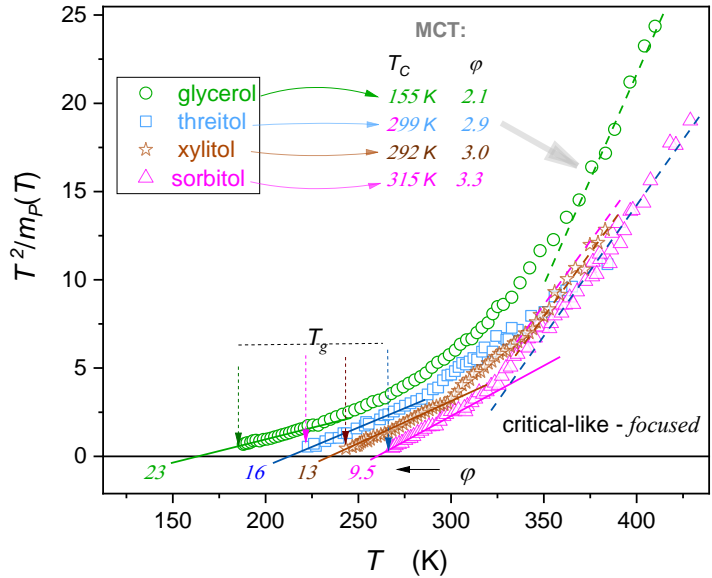

*Fig. 7. Distortion-sensitive plot focused on searching domains, preferably described by the critical-like description Eq. (29): straight lines indicate them.*

Figures 6 and 7 present results of linearized, derivative-based analysis focused on VFT (Eq. (3)) and the critical-like (Eq. (27)) scaling of previtreous dynamics. Distortion-sensitive analysis reveals that near $T_g$ VFT portrayal obeys only for glycerol. Such a behavior ceases to be optimal when shifting from glycerol to sorbitol in a homologous series, as visible by emerging non-linearity. A reversed behavior occurs when testing critical-like portrayal preference (Eq. (27)). It is optimal for sorbitol and becomes non-optimal when shifting from sorbitol to glycerol in the tested series, as visible in Fig. 7. Notable that for sorbitol, which molecule shows the 'strongest' uniaxial features, the exponent $\varphi \approx 9.5$ $T \to T_g$. Such a value is approximately the same as in rod-like liquid crystalline glass-formers [105-109] and roughly the same as introduced by Colby within the Dynamical Scaling Model [56, 57]. In high-temperature domain, well above $T_g$, second critical-like domain emerges (Fig. 7), in agreement with the MCT approach expectations for HT ergodic part of previtreous domain.

Plot shows the values of MCT [49, 83] exponents 'critical' temperatures in high-temperature dynamic domain. Regarding low-temperature domain near $T_g$ the following parameters have been obtained: (i) for glycerol $T_C = 155K$ and $\phi \approx 23$, (ii) $T_C = 214K$, $\phi \approx 16$ for threitol, (iii) for xylitol $T_C = 233K$ and $\phi \approx 13$, (iv) $T_C = 254K$, $\phi \approx 9.5$ for sorbitol. The latter extends up to ca. $T_g + 50K$.

Recently, one of the authors (ADR) showed the common pattern empirically for the evolution of the apparent fragility for ten glass formers, covering low molecular weight liquids, liquid crystals, plastic crystals, polymers, and resins [86]:

$$m_P(T) = \frac{d \log_{10} \tau(T)}{d(T_g/T)}\bigg|_{T>T_g} = \frac{A}{T-T_g^*} \tag{30}$$

The extrapolated singular temperature $T_g^* < T_g$ may be easily determined from the condition $1/m_P(T_g^*) = 0$. Linking the above empirical equation with definition of apparent fragility, one obtains the differential equation, which solution leads to the following scaling dependence for the previtreous behavior [86]:

$$\tau(T) = C_\Gamma \left(\frac{T-T_g^*}{T}\right)^{-\Gamma} \left[\exp\left(\frac{T-T_g^*}{T}\right)\right]^\Gamma = C_\Gamma (t^{-1} \exp t)^\Gamma \tag{31}$$

where $t = (T - T_g^*)/T$

The power exponent can be expressed via basic empirical metrics of the glass transition: $\Gamma = m \ln 10 \, (T_g/T_g^*)/(1/(\Delta T_g^*/T_g) - 1)$, $\Delta T_g^* = T_g - T_g^*$. The unique feature of Eq. (31) is the 'activation-critical' (AC) formula linking critical-like and activation (SA-type) features. The value of the exponent determines their relative share in the previtreous effect. The parallel relation may be introduced for the high-temperature dynamic domain. In this case, the singular temperature $T_B^* < T_B$ and the power exponent $\Gamma_B$ replace parameters in Eqs. (31).

**Table I** Values of parameters for the AC Eq. (31). $T_g^*$, $\Gamma$, $C_\Gamma$ are for LT dynamic domain. Parameters for the HT dynamic domain are denoted as $T_B^*$, $\Gamma_B$, $C_{\Gamma B}$ and related numbers are in italic.

| Glass-former | $T_g$ (K) | $T_g^*$ (K) $T_B^*$ | $\Gamma$ | $\log_{10} C_\Gamma$ $\log_{10} C_{\Gamma B}$ |
|---|---|---|---|---|
| Glycerol | 186.0 | 147.1 | 34.3 | -19.48 |
|  |  | *225.1* | *7.05* | *-13.31* |
| Threitol | 224.2 | 200.4 | 19.3 | -17.47 |
|  |  | *270.1* | *4.7* | *-12.6* |
| Xylitol | 247.6 | 231.5 | 16.02 | -17.08 |
|  |  | *299.3* | *4.1* | *-12.6* |
| Sorbitol | 267.0 | 259.1 | 11.4 | -15.33 |
|  |  | *300.1* | *3.85* | *-12.7* |

Emerging from comparison of Figs. 6 and 7, interplays between activation-type (SA) and critical-like (CL) dynamics indicate 'mixed' scaling as a possible optimal parameterization for the homologous series of tested polyols. Such a relation (Eq. (31)) has been introduced recently and validated for a set of glass-forming systems [86].

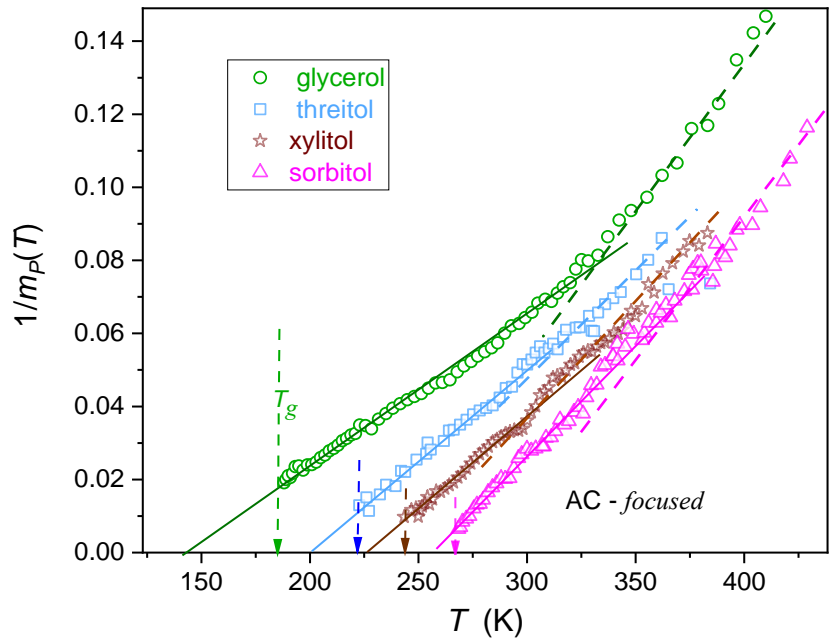

*Fig. 8. Previtreous universal changes of the apparent fragility in the previtreous domain of polyols emerging due to for $1/m(T)$ vs. T plot. Arrows indicate glass temperatures. Linear behavior is linked to Eq. (30).*

Values of parameters are collected in Table I. It is worth noting, that values of singular temperatures $T_g^*$ and $T_B^*$ may be easily determined prior to fitting of $\tau(T)$ experimental data by analysis, which results are shown in Fig. 8. This causes final fit of $\tau(T)$ can be limited only to two parameters. Values of parameters given in Table I show that the power exponent in Eq. (31) is responsible for the relative impact of the critical-like and activation contributions to previtreous effect.

As mentioned above, one of solutions to avoid problems with the reliability of VFT portrayal was supplementation by power exponent, as in Eq. (14). For such relation, one can consider the following transformation:

$$ln\tau(T) = ln\tau_\infty + \left(\frac{\Phi}{T-T_0}\right)^w \qquad \frac{dln\tau}{dT} = w\Phi^w \left(\frac{1}{T-T_0}\right)^{w-1} \qquad (32)$$

Linearized analysis limited to first derivative of experimental data is possible for the assumed model-related value of the exponent. The leading position seems to play the Bendler-Schlezinger model [79] for which $w = 3/2$. For such a parameterization following linearized and distortion-sensitive analysis may be considered:

$$\frac{dln\tau}{dT} = (1.5\Phi^{1.5})(T-T_0)^{-1/2} \quad \Rightarrow$$

$$\Rightarrow \quad \left(\frac{dln\tau}{dT}\right)^{-2} = (2.25\Phi^{-3})T - (2.25\Phi^{-3})T_o = AT + B \qquad (33)$$

However, the analysis based on Eq. (33) does not show linear behavior in ultraviscous domain for any member of polyols, which puts applicability of Eq. (14) for real systems in question.

## 5. Activation energy index for analysis previtreous dynamics

Hecksher et al. [44] proposed to focus on activation energy index introduced by Dyre and Olsen [43]: $I_{DO}(T) = - d\ln E_a(T)/d\ln T = (dE_a/E_a)/(dT/T)$, i.e., to transform experimental data $\tau(T) \rightarrow I_{DO}(T)$. The required apparent activation energy was calculated from general SA Eq. (1), $E_a(T) = RT \ln(\tau(T)/\tau_\infty)$ assuming universal value of $\tau_\infty$. In ref. [44] the analysis of 42 low-molecular-weight glass formers led to the conclusion: '*…there is no compelling evidence for the Vogel–Fulcher–Tammann (VFT) prediction that the relaxation time diverges at a finite temperature. We conclude that theories with a dynamic divergence of the VFT form lack a direct experimental basis*.'

It was formulated by comparing experimental $I_{DO}(T)$ evolutions with model $I_{DO}(T)$ dependences for the VFT relation and two proposed functions without finite temperature singularities (denoted as *FF1* and *FF2* in ref. [44]).

In subsequent years, ref. [44] has become an inspiration for developing theoretical models avoiding finite temperature singularities below $T_g$. However, the pre-exponential factor assumption regarding universal, constant value $\tau_\infty = 10^{-14} s$ in ref. [44], poorly correlate with experimental evidence and may lead to a bias for calculated values of $E_a(T)$ and then $I_{DO}(T)$. In ref. [45] protocol avoiding this problem was proposed: apparent activation energy was calculated as a solution of differential equation resulting from SA Eq. (1) or the selected set of $\tau(T)$ experimental data:

$$R \frac{d \ln \tau(T)}{d(1/T)} = \frac{1}{T} \frac{dE_a(T)}{d(1/T)} + E_a(T) \tag{34}$$

In Eq. (34) $d \ln \tau / d(1/T) = H_a^{'}(T) = H_a(T)/R$, where $H_a(T)$ is apparent activation enthalpy. As results from Eq. (34) $H_a(T) \neq E_a(T)$ for the SA dynamics. In refs. [45-47] the analysis exploring Eq. (34) for determining the apparent activation energy was applied for 26 glass formers, ranging from low-molecular-weight liquids, polymers, plastic crystals to liquid crystals. The common 'universal' pattern of the index was found: $1/I_{DO} = a + bT$. Subsequently, a hypothetical general formula for activation energy index was found [46, 47]:

$$I_{DO}(T) = nT_0/(T - T_0) = \frac{nT_0}{T - T_0} \tag{35}$$

It was concluded from $I_{DO}(T)$ derived form for VFT, Avramov-Milchev (AM), MYEGA, and critical-like (CL) dependences [45]. The study of experimental data in ref. [61] showed that $0.18 < (n = -1/a) < 2.2$. VFT portrayal appears for systems characterized by $n = 1$ with orientational, uniaxial ordering, whereas $n \sim 0.18$ for systems with translational symmetry. Notably, that for MYEGA [45] equation $a = 0$, and for the Avramov-Milchev (AM) [45] dependence $b = 0$ and then $1/I_{DO}(T) = const$. For VFT and CL scaling relations: $a \neq 0$ and $b \neq 0$.

Table II  Basic model relation and related forms of reciprocal of the activation energy index, including the corresponding parameter '$n$'. The experimental evidence is also summarized.

| | Model Equation | $1/I_{DO}$ | Parameter $n$ |
|---|---|---|---|
| **Theory/Model** | VFT: $\tau(T) = \tau_\infty \exp\left(\frac{D_T T_0}{T-T_0}\right)$ | $\left(\frac{1}{T_0}\right)T - 1$ | 1 |
| | MYEGA: $\tau(T) = \tau_\infty \exp\left[\frac{C}{T}\exp\left(\frac{K}{T}\right)\right]$ | $\left(\frac{1}{C}\right)T$ | 0 |
| | AM: $\tau(T) = \tau_\infty \exp\left(\frac{A_{AM}}{T^D}\right)$ | $\frac{1}{D-1}$ | undefined |
| | Critical-like: $\tau(T) = \tau_\infty(T - T_C)^{-\varphi}$ | $\left(\frac{1}{\varphi}\right)T - \frac{T_C}{\varphi}$ | ~ 0.2 (PC)<br>~1.5 (LC) |
| | CGE: $\tau(T) = \tau_\infty \exp\left[C\left(\frac{1}{T} - \frac{1}{T_0}\right)^2\right]$ | $\frac{T - T_0}{T + T_0}$ | undefined |
| Experiment: | $1/I_{DO} = aT + b$ , $n = -1/b$ | | |
| (PC: *critical-like*) | $\Leftarrow 0.18 < n < 2.2 \Rightarrow$ (LC: *critical-like*) | | |
| | $n = 1$ (SA: VFT) | | |

Elmatad et all. introduce a parabolic scaling plot of activation energy defining a cross-over temperature $T_o > T_m$ and $J$ is a parameter setting the excitation energy [59]. The activation energy for this scaling approach can be written as:

$$E_a(T) = \left(\frac{J}{T_o}\right)^2 T \left(\frac{T_o}{T} - 1\right)^2 \quad , \text{ for } T < T_o \tag{36}$$

Consequently, following formula for activation energy index is obtained:

$$I_{DO}(T) = \frac{T + T_o}{T - T_o} \tag{37}$$

Reciprocal of the index does not follow linear behavior and exhibits an artificial anomaly associated with onset temperature $T_o$, related to crossover between Arrhenius and Super-Arrhenius dynamics domains in high-temperature region.

The relation linking apparent activation energy, activation energy index and configurational entropy derived in Refs. [45, 107] is notable:

$$I_{DO}(T) = \frac{1}{T}\frac{d\ln E_a}{d(1/T)} = -\frac{1}{TS_C(T)}\frac{dS_C(T)}{d(1/T)} \tag{38}$$

By applying the experimental evidence for apparent activation energy index evolution (Eq. (35)) one obtains following relation for previtreous changes of configurational entropy [45]:

$$S_C(T) = S_0 \left(1 - \frac{T_0}{T}\right)^n = S_0 t^n \qquad (39)$$

where $t = (T - T_0)/T$.

It is worth stressing, that essential difference between Eq. (39) and classic relation for configurational entropy (Eq. (9), which can be retrieved for $n = 1$ and $T_0 = T_K$. Equation (39) is associated with a power exponent, for which experimental evidence indicates a link to local symmetry of glass-former. Moreover, behavior described by Eq. (40) extends in low-temperature (*ultraviscous, ultraslowed*) dynamical domain extending up to $\sim T_g + 80K$. All these show a notable similarity to critical phenomena. Substituting Eq. (39) to the basic AG model relation (Eq. (7)), one obtains the following dependence:

$$\tau(T) = \tau_\infty exp\left(\frac{DT^{n-1}}{(T-T_0)^n}\right) = \tau_\infty exp\left(\frac{D/T}{t^n}\right) \qquad (40)$$

where $D = A\Delta\mu/S_0$.

For $n = 1$ Eq. (40) converts into basic VFT Eq. (3), and then $D = D_T$, $T_0$ is basic VFT singular temperature. Extended VFT Eq. (40) with $n \neq 1$ was independently introduced to portray previtreous dynamics in polyvinylidene disulfide (PVDF) and BST ferroelectric microparticles [110] as well as for relaxor ceramics [111]. Formally, Eq. (40) is associated with four fitted parameters, but values of $n$ and $T_0$ may be estimated from activation energy index analysis or heat capacity data. It is worth stressing, that emerging similarity of Eq. (40) and RFOT model [112] general relation (Eq. 10). The latter can be retrieved if the power exponent $n = \alpha = \psi/(d - \theta)$.

It is notable, that Eq. (40) for $\tau(T)$ evolution was obtained assuming configurational entropy derived from $\tau(T)$ experimental data and expressed by Eq. (39). Crucial validation requires obtaining behavior described by Eq. (39) from thermodynamic data analysis. At first sight, credibility of Eq. (39) seems to be doubtful since experimental confirmation of classic Eq. (9), linked to $n = 1$, is very extensive [26-31, 38, 49, 54, 75]. Notwithstanding, analysis based on non-linear fitting of experimental data remote from singular Kauzmann temperature $T_K$. The very recent report [113] cope with this essential feature of previtreous behavior for glass transitions, analyzing high-resolution experimental data for 8 glass-forming systems via the following distortions-sensitive approach for Eq. (39):

$$lnS_C(T) = lnS_0 + nln\left(1 - \frac{T_K}{T}\right) \Rightarrow \frac{dlnS_C}{d(1/T)} = \frac{nT_K}{1 - T_K/T} \qquad (41)$$

Consequently, one obtains the following linear behavior validating Eq. (39) for the plot defined by following relations:

$$\left[\frac{dlnS_C(T)}{d(1/T)}\right]^{-1} = \frac{1}{nT_K} - \left(\frac{1}{n}\right)\left(\frac{1}{T}\right) = A + B\left(\frac{1}{T}\right) \qquad (42)$$

Table III Values of parameter n reported in ref. [113] using the derivative-based analysis via Eq. (42).

| **System** | Sorbitol | 8*OCB (LC: rod-like) | Ethanol | Glycerol | Diethyl phtalate | Cycloheptanol (PC: ODIC) |
|---|---|---|---|---|---|---|
| ***n*** | 1.57 | 1.51 | 1.28 | 1.04 | 0.98 | 0.18 |

Figures 9 and 10 show the results of such analysis for glycerol and propanol. Table III below presents the summary of results discussed in ref. [113]. These results support the generalized Eq. (39) for the configurational entropy, with the exponent $n \neq 1$, within frames indicated by ref. [113].

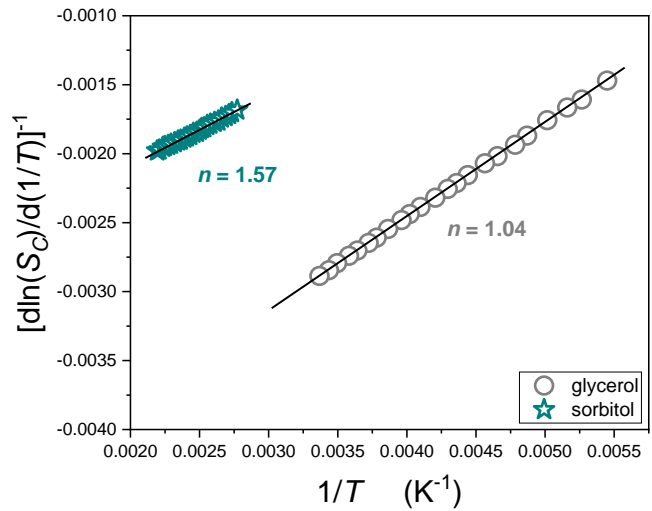

*Fig. 9. The derivative-based analysis of configurational entropy (Eq. (42)) for two selected supercooled liquids focused on testing new relation $S_C(T)$ behavior given by Eq. (41).*

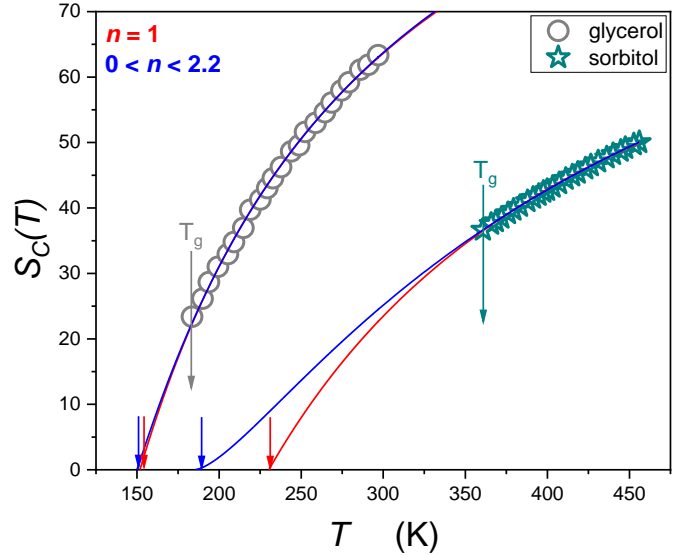

*Fig. 10. Temperature dependences of configurational entropy in glycerol and sorbitol and their portrayals via Eq. (39), with parameters derived using the analysis presented in Fig. 9. In red, the description by the n =1 is shown.*

Figure 10 shows that direct portrayals of $S_C(T)$ experimental data for $T > T_g$ via discussed dependences, associated with $n \neq 1$, and $n = 1$, yields almost indistinguishable fitting qualities, particularly when including impact of the experimental error. Notwithstanding, generalized relation for configurational entropy evolution (Eq. (39)) yield notable correction in estimation of the Kauzmann temperature, as shown in Fig. 10 and ref. [113].

As a general comment for the analysis exploring activation energy index for relaxation, one should note that it applies the second and even the third-order derivative of $\tau(T)$ experimental data, supported by numerical filtering using the Savitzky-Golay principle [45]. All these introduce some degree of uncertainty to results.

## 6. Distortion-sensitive tests of previtreous dynamics under pressure

Isothermal compressing constitutes an alternative way of approaching glass transition. Relation to describe pressure-viscosity behavior was firsty proposed by Barus (B) in 1893, via the relation $\eta(P) \propto \exp(\alpha P)$, $\alpha = const$ [114]. Nowadays, this relation is used in the super-Barus (SB) form, with pressure-dependent coefficient $\alpha(P)$:

$$\tau(P) = \tau_\infty^T \exp(\alpha(P) \times P) = \tau_\infty^T \exp\left(\frac{V_a(P)}{RT} P\right) \tag{43}$$

where $T = const$, $P < P_g$ and $P_g$ is for glass (vitrification) pressure; $V_a(P)$ denotes apparent activation volume.

Note, that in both Barus (B) and Super-Barus (SB) equations, the pre-exponential factor:

$$\tau_\infty^T = \tau(P = 0) \approx \tau(P = 0.1 MPa) \tag{44}$$

Williams introduced the activation volume into this relation in 1964 [68, 69], as follows:

$$V_a = RT \frac{d \ln \tau(P)}{dP}, \tag{45}$$

where $V_a(P) = V_a = const$ in the given pressure domain, which is related to the basic Barus relation. The latter equation is not valid for pressure-dependent apparent activation volume, i.e. for the SB dynamics:

$$RT \frac{d \ln \tau(P)}{dP} = V_a(P) + P \frac{dV_a}{dP} \tag{46a}$$

and consequently

$$V^\#(P) = RT \frac{d \ln \tau(P)}{dP} \neq V_a \tag{46b}$$

Taking into account definition of pressure-related apparent fragility (steepness index) defined below, one may show that $V^\#(T) \propto m_T(P)$. The general Super-Arrhenius as well as Super-Barus relation may be obtained by linking Eqs. (1) and (44) [115]:

$$\tau(T, P) = \tau(T)\tau(P) = \tau_{ref.} \exp\left(\frac{E_a(T) + PV_a(P)}{RT}\right) \tag{47}$$

where $T > T_g$ and $P < P_g$. A similar relation occurs for $\eta(T, P)$ changes.

Following eq. (47), an isobaric temperature evolution is described by:

$$\tau(T) = \left[\tau_\infty^T \exp\left(\frac{PV_a(P)}{RT}\right)\right] \exp\left(\frac{E_a(T)}{RT}\right) = \tau_\infty^P \exp\left(\frac{E_a^P(T)}{RT}\right) \tag{48}$$

Eq. (48) correlates with SA Eq. (1) for $P = 0$. Such a isobar may be approximated by temperature studies under atmospheric pressure ($P \approx 0.1 MPa$). Note, that Eq. (48) shows a pattern of changing in pre-exponential factor, when carrying out temperature tests under higher pressures.

For isothermal pressure-related previtreous behavior, plot $log_{10}\tau(P)$ or $log_{10}\eta(P)$ vs. $P/P_g$ is often indicated as a possible pressure counterpart of the Angell plot. However, such a presentation of data leads to a gamut of curves instead of a compact representation characterizing temperature-related Angell plot. The mentioned normalized $\tau(P)$ or $\eta(P)$a evolution lead to folloing pressure-related steepness index (apparent fragility) [116, 117]:

$$m_T(P) = \frac{dlog_{10}\tau(P)}{d(P/P_g)} \quad , \quad m_T = m_T(P \to P_g) \tag{49}$$

where $m_T$ denotes pressure-related fragility metric.

The above yields $\mu_P = log_{10}\tau(P_g) - log_{10}\tau_\infty^P$ for the minimal fragility characterizing basic Barus dynamics: fragility. However, depending on tested isotherm, it may range from 1 to even 14. It is also notable, that Eqs. (48) and (49) lead to the 'artificial' anomaly of apparent fragility for $P \to 0$ [116, 117]. All above indicates significant inconsistencies for general characterization of previtreous effect in super-pressed liquids.

Similar to basic SA Eq. (1) and SB Eq. (43) does not enable portrayal of experimental data due to unknown form of an evolution of apparent activation volume. Consequently, replacement relations are necessary. In 1972 Johari and Whalley (JW) applied the following empirical dependence for portraying experimental data in super-pressed glycerol ($T = 20\,^oC$) [118]:

$$\tau(P) = \tau_\infty^T exp\left(\frac{J}{P_0-P}\right) \tag{50}$$

where $J = const$, $P < P_g$, extrapolated singular pressure $P_0 > P_g$.

However, this relation may reliably portray experimental data only if they are relatively close to basic Arrhenius/Barus pattern (i.e. 'strong' glass-formers) or in a 'narrow' range of pressures. Moreover, it cannot be reduced to basic Barus equations with $V_a(P) = V_a = const$, what is an essential feature required for any SB replacement scaling relation. When comparing general SB Eq. (43) and JW Eq. (50) a significant inconsistency occurs also for the pre-exponential factor: in Eq. (50) $\tau_\infty^T \neq \tau(P = 0)$.

In 1998, an application of new BDS facilities and designs of measurement capacitors placed within pressure chambers enabled obtaining the high-resolution $\tau(P)$ experimental data for ultraviscous glycerol compressed up to 0.35 GPa for $T \sim 260K$, showing the explicit fragile behavior. Analysis showed a limited adequacy of Eq. (50) and the fair portrayal by an empirical relation [119]:

$$\tau(P) = \tau_\infty^T exp\left(\frac{J(P)}{P_0 - P}\right) = \tau_\infty^T exp\left(\frac{D_P P}{P_0 - P}\right) \tag{51}$$

It can fair portray dynamics for both 'strong' and 'fragile' glass-formers. It also introduced fragility strength coefficient $D_P$ for a pressure path. Notably, it may be reduced to basic Barus equation and the prefactor $\tau_\infty^T = \tau(P = 0)$, as in basic Barus and Super-Barus relations (Eq. 43). It also can be derived from the VFT Eq. (3) by a simple substitution $T = \delta/P$ and $\delta = const$, i.e., basic qualitative link between cooling and compressing:

$$\tau(T) = \tau_\infty exp\left(\frac{D_T T_0}{T - T_0}\right) \quad \Rightarrow \tag{52}$$

$$\tau_\infty^T exp\left[\frac{D_T(\delta/P_0)}{\delta/P - \delta/P_0}\right] = \tau_\infty^T exp\left[\frac{D_T(\delta/P_0)}{\delta(1/P - 1/P_0)}\right] = \tau_\infty^T exp\left[\frac{D_T(\delta/P_0)}{\delta\left(\frac{P_0 - P}{P_0 P}\right)}\right] = \tau_\infty^T exp\left[\frac{D_T P}{P_0 - P}\right]$$

Notwithstanding, in a new PVFT Eq. (51), problem of pre-exponential factor inconsistency, characterizing SB Eq. (43), remains. It may be solved, taking into account, that liquids or solids can be isotopically stretched, which is equivalent to negative pressures and passing $P = 0$ without any hallmark [120, 121]. The stretching is possible until an absolute stability limit spinodal $P_{SL} < 0$, where intermolecular interactions break, is reached. Experimental evidence of smooth passing from the 'positive' to the 'negative' pressures domains in glass-forming liquids was shown by Angell and Quing [122]. All these led to generalized SB and PVFT relations [95]:

$$\tau(P) = \tau_\infty^T \exp\left(\frac{V_a(P)}{RT}\Delta P\right) \tag{53a}$$

$$\tau(P) = \tau_\infty^T \exp\left(\frac{D_P^{SL}(P-P_{SL})}{P-P_0}\right) = \tau_\infty^T \exp\left(\frac{D_P^{SL}\Delta P}{P-P_0}\right) \tag{53b}$$

where $D_P^{SL}$ is fragility strength corrected by impact of stability limit (*SL*) pressure and $\Delta P = P - P_{SL}$.

Note, that Eq. (53a) resemble the one proposed by Kießkalt yet in 1927 [127]: $\eta = \eta_o e^{a(P-P_0)}$. but $P_0$ was referred to some 'characteristic positive' pressure and negative pressures domain was not considered there.

Eqs. (53a) and (53b) directly lead to plot $log_{10}\tau(P)$ vs. $\Delta P = P - P_{SL}$ as pressure counterpart of Angell plot, see ref. [95]. It is worth noting, that Eqs. (53a and 53b) are associated with similar values of pre-exponential factor $\tau_\infty^T \neq \tau(P_{SL}) \sim 10^{-11} s$, yielding one value for minimal (Barus-related) fragility $\mu_P \approx 13$, for arbitrary tested isotherm. All these led to a new definition of apparent fragility $m_T^{\Delta P}$, which can be linked to the 'old' one as follows [95]:

$$m_T^{\Delta P} = \frac{d\log_{10}\tau(P)}{d(\Delta P/\Delta P_g)} = \Delta P_g \frac{d\log_{10}\tau(P)}{d(\Delta P)} = \frac{\Delta P_g}{P_g}\frac{d\log_{10}\tau(P)}{d(P/P_g)} = \frac{\Delta P_g}{P_g} m_T(P) \tag{54}$$

where $\Delta P = P - P_{SL}$, $\Delta P_g = P_g - P_{SL}$ and $P \leq P_g$

The new PVFT Eq. (53b) contains four adjustable parameters. However, their validity may be tested by derivative-based analysis given below, which also yields optimal values of basic parameters [87, 95]:

$$\left[\frac{d\ln\tau(P)}{dP}\right]^{-1/2} = (D_P P_0)^{-1/2} P_0 - (D_P P)^{-1/2} P = A + BP \tag{56}$$

$$\left[\frac{d\ln\tau(P)}{dP}\right]^{-1/2} = \left[D_P^{SL}(P_0 - P_{SL})\right]^{-1/2} P_0 - \left[D_P^{SL}(P_0 - P_{SL})\right]^{-1/2} P = A + BP \tag{57}$$

Eqs. (56) and (57) are for basic PVFT Eq. (51) and new N-PVFT Eq. (53b), respectively. The comparison of Eqs. (56) and (57) show a link between fragility strength in Eqs. (51) and (53b) [87]: $D_P^{SL} = D_P[(P_0 - P_{SL})/P_0]$.

Despite success of PVFT relation, there is still no theoretical models offering its derivation. Notwithstanding, one may transform the basic VFT Eq. (3) into the PVFT Eq. (53) by a simple substitution $T = A/P$: $\tau(T) = \tau_\infty(T)\exp(D_T T_0/(T-T_0))$ $\Rightarrow$ $\tau(P) = \tau_\infty \exp[D_T(A/P_0)/((A/P) - (A/P_0))] = \tau_\infty \exp[D_T P/(P_0 - P)]$. Such a simple link leads to a question of whether significant problems of basic VFT relation also extend to PVFT one? Notably, relation resembling basic PVFT Eq. (51) was reported in 1963 by Roelands et al. (1963): $\eta(P) = \eta_0 \exp(\alpha_0 P/(1 + R_3 P))$, where $R_3$ and $\alpha_0$ are constants, for describing viscosity changes in lubricating oils [123]. One can also consider yet another relation, introduced by Roeland, originally for viscosity [123]:

$$\tau(P) = \tau_\infty \exp(R_1 P^{R2}) \tag{58}$$

were $R_1$ and $R_2$ are system-dependent empirical constants.

Although the relation was developed by heuristic considerations. It can also be directly derived from the Avramov-Milchev Eq. (19), if taking into account basic relationship between cooling and compressing, $T = A/P$:

$$\tau(T) = \tau_\infty \exp\left(\frac{A_M}{T^D}\right) \Rightarrow T = \frac{A}{P} \Rightarrow \tau(P) = \tau_\infty \exp(A'_M P^D) \tag{59}$$

It shows that $R_2$ parameter in Roelands Eq. (58) reflects a pressure-related fragility. Note, that Eqs. (58) and (59) suffer from the same problem with pre-exponential factor values like basic PVFT Eq. (51). This problem disappears, if an 'extended version' of Eqs. (58) and (59) is considered:

$$\tau(P) = \tau_\infty \exp\left(R'_1 (P - P_{SL})^{R'_2}\right) = \tau_\infty \exp\left(R'_1 (\Delta P)^{R'_2}\right) \tag{60}$$

where the absolute stability limit pressure $P_{SL} < 0$.

For Eqs. (58) and (60) following distortions-sensitive test may be proposed:

$$ln\,\tau(P) = ln\,\tau_\infty + R'_1(\Delta P)^{R'_2} \Rightarrow \frac{dln\,\tau(P)}{dP} = R'_1 R'_2 (\Delta P)^{R'_2 - 1} \Rightarrow ln\left(\frac{dln\,\tau(P)}{dP}\right) = lnV^{\#}(P) = R'_1 R'_2 (R'_2 - 1)ln(\Delta P) \tag{61}$$

In basic Roeland's Eq. (58) is simplified to following dependence:

$$ln\left(\frac{dln\,\tau(P)}{dP}\right) = lnP \tag{62}$$

where $R = R'_1 R'_2 (R'_2 - 1)$.

Eqs. (58) and (60) are validated if the linear domain appears in the plot $lnV^{\#}(P)$ vs. $P$ or $\Delta P$. Subsequent linear regression fit may yield optimal values of basic parameters. The above reasoning can be implemented for introducing other relations describing pressure-related SB dynamics. Taking MYEGA Eq. (12) as a reference, one may derive a following dependence:

$$\tau(T) = \tau_\infty \exp\left[\frac{C}{T}\exp\left(\frac{K}{T}\right)\right] \Rightarrow T = \frac{A}{P} \Rightarrow \tau(P) = \tau_\infty \exp(C'P \exp(K'P)) \tag{63}$$

where, $C' = C/A$ and $K' = K/A$. Its application can be validated by appearance of linear domain in plot defined by following relation:

$$\ln \tau(P) = \ln \tau_\infty + C'P \exp(K'P) \Rightarrow V^\# = \frac{\ln \tau(P)}{dP} = C' \exp(K'P)[K'P + 1] \Rightarrow$$

$$\ln V^\# = \ln C' + K'P + \ln(K'P + 1) \approx 2K'P + \ln C' = aP + b \tag{64}$$

where the term $\ln(K'P + 1)$ is extened in the Taylor series.

The question arises whether Eq. (63) to be a kind of 'unicorn' with a definite advantage over PVFT or Roland relations, as is the case for MYEGA relation in relation to VFT and Avramov ones for temperature evolution of relaxation time or viscosity?

One may also consider a pressure counterpart of RFOT-developed AG model relation:

$$\tau(T) = \tau_\infty \exp\left(\frac{A_G}{T(S_C(T))^\alpha}\right) \Rightarrow \tau(P) = \tau_\infty \exp\left(A'_G P (S_C(P))^\alpha\right) \tag{65}$$

where basic AG model is related to the exponent $\alpha = 1$.

Following Eq. (65) one can propose the relation for pressure evolution of the configurational entropy:

$$S_C(P) = \frac{S_0}{(P_0 - P)^{n'}} \tag{66}$$

where the exponent $n'$ may be related to both empirical symmetry-related exponent $n$ and the RFOT exponent $\alpha$.

For experimental validation of Eq. (66) necessary are challenging and still non-available pressure-related changes in configurational entropy.

Recently, it has been discovered that transforming $\tau(P)$ experimental data to pressure-related steepness index, i.e., apparent fragility 'universal' dependence for plot $[m_T(P)]^{-1}$ vs. $P$. This directly yields to following relation [128]:

$$\frac{1}{m_T(P)} = a_{HP} + b_{HP}P \quad \rightarrow \quad m_T(P) = \frac{A_{HP}}{P^* - P} \tag{67}$$

where $P_B^{1/m} < P < P_g$ and $P^* > P_g$; singular pressure is estimated via the condition $1/m_T(P_g^*) = 0$.

Linking Eq. (67) with definition of pressure-related apparent fragility, one obtains a differential equation, which solution leads to new critical-like relation for portraying pressure-related previtreous dynamics [128]:

$$\tau(P) = \tau_{\infty P}(P^* - P)^{-\Psi} \qquad (68)$$

Using the preliminary analysis byt Eq. (67) one may determine a singular pressure $1/m_T(P^*) = 0$ or $1/V^\#(P) = 0$ and then 'discontinuity': $\Delta P_g^* = P_g^* - P_g$. All these allow estimating the power exponent in Eq. 568: $\Psi = ln10 \left( \Delta P_g^* / P_g^* \right) m_T(P_g)$.

The above discussion of Super-Barus dynamics has been tested using experimental the Author's experimental data for glycerol and xylitol. Note, that such results are still hardly evidenced, particularly when considering GPa domain. First challenging problem for studies under high pressure of liquid system is an isolation of tested samples from pressurized medium.

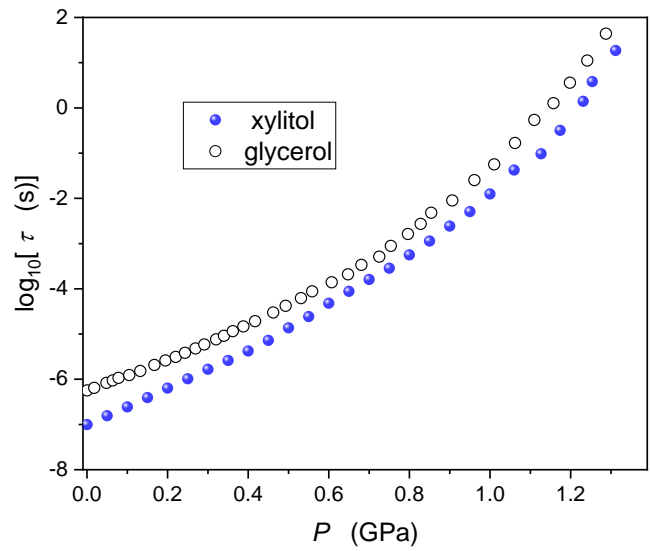

*Fig. 11. Pressure evolution of primary relaxation time in super-pressed xylitol (T = 280 K isotherm) and glycerol (T = 250 K isotherm).*

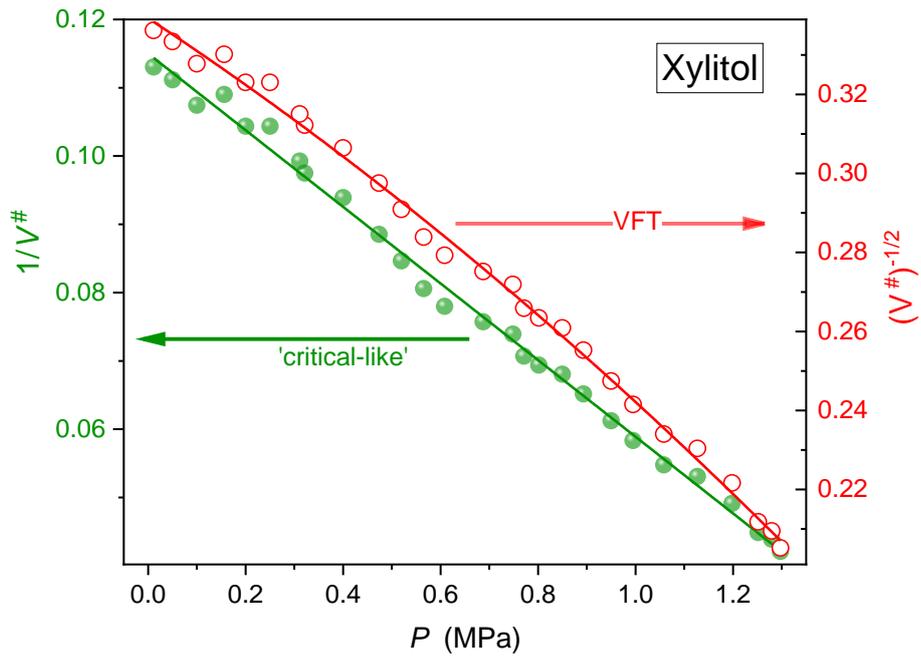

*Fig. 12. Linearized and derivative-based focused testing for PVFT (Eqs. 51, 56) and critical-like portrayals (Eqs. 67, 68) in super-pressed xylitol, based on experimental data shown in Fig.11. Mentioned scaling relations are validated by the linear behavior's emergence, which occurs only for the critical-like portrayal.*

High-pressure BDS studies are still limited to frequency range $f \sim 10 MHz$, i.e., relaxation time $\tau < 10^{-7} s$. It means, that experimentally only ultraviscous/ultraslowing domain near glass pressure is available.

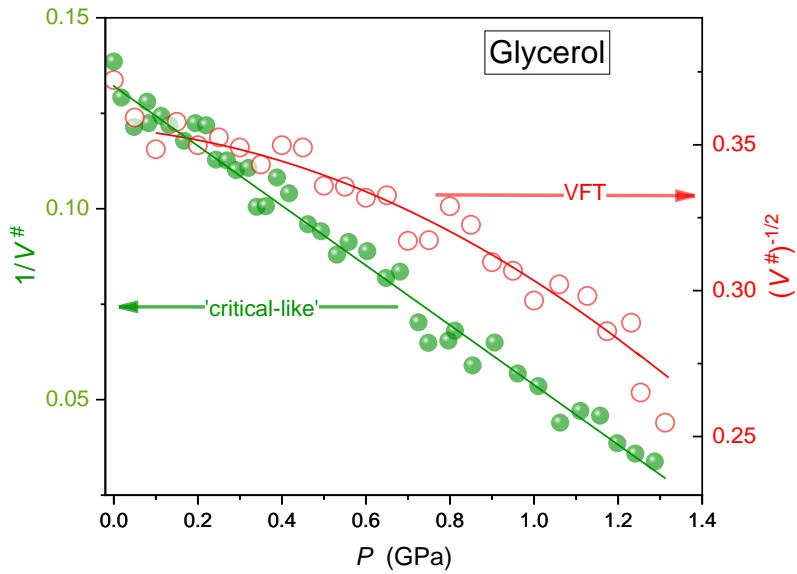

*Fig. 13. Linearized and derivative-based focused on testing PVFT (Eqs. 51, 56) and critical-like portrayals (Eqs. 67, 68) in super-pressed glycerol, based on experimental data shown in Fig.11. Mentioned scaling relations are validated by emergence of linear behavior, what occurs only for critical-like portrayal.*

Figures 12 and 13 shows the comparison of the PVFT (Eqs. (51) and (56), in red) and the critical-like (Eqs. (68) and (67)) via the linearized and distortions-sensitive analysis for supercooled glycerol and xylitol. For both cases, the clear prevalence for the critical-like portrayal is manifested via the explicit linear behavior.

Figure 14 presents the linearized, derivative-based test results for the unicorn Eq. (63), by coupled Eq. (64). Results of analysis that it may be validated in the range of pressure ~ 1 GPa, much broad range of pressure than for standard description via PVFT Eq. (61).

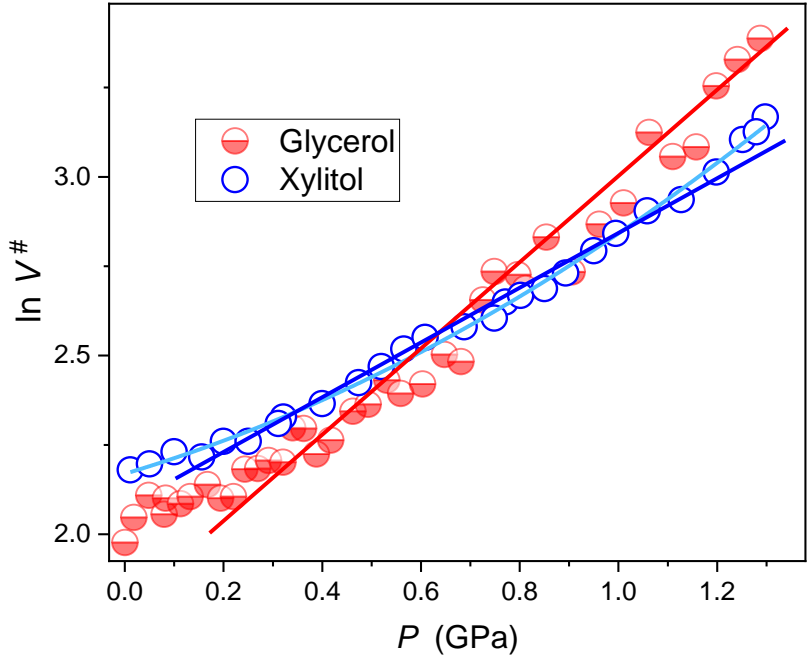

*Fig. 14. Linearized and derivative-based analysis tested validity of portrayal of τ(P) experimental data shown in Fig. 11, using the UNICORN Eq. (63), indicated by the emergence of linear behavior (see Eq. (64)). The light blue line shows preference for second-order polynomial parameterization in a total range of pressures.*

As mentioned in this section, estimating evolution of apparent activation energy hypothetically responsible for previtreous slowing down requires solution of a differential equation associated with the second-order derivative of experimental data. For pressure path, leading role plays Barus-Williams Eq. (43) governed by apparent activation volume. Most often, it is determined as $V^{\#}(P) = dln\tau(P)/dP$. However, such an estimation for SB behavior is incorrect, namely based on SB Eq. (43) one obtains [129]:

$$V^{\#}(P) = \frac{dln\tau(P)}{dP} = \left(\frac{1}{RT}\right) V_a(P) + \left(\frac{1}{RT}\right) P \frac{dV_a(P)}{dP} , \quad \text{for } T = const \qquad (69)$$

The above relation clearly shows that $V^{\#}(P) \neq V_a(P)$, except the case $P \to 0$ or for basic Barus behavior. As shown in ref. [129] for SB dynamics $V^{\#}(P) \propto m_T(P)$. However, for pressure-related previtreous effects, pre-exponential factor is perfectly known what allows to calculate the real apparent activation volume directly from the SB Eq. (43) [129]:

$$V_a(P) = \frac{RT}{P} ln\left(\frac{\tau(P)}{\tau_\infty}\right) \quad , \quad T = const \tag{70}$$

In ref. [129] also the relation for pressure evolution of apparent activation volume was derived:

$$V_a(P) = \frac{C}{\Delta P} + \frac{\Psi}{\Delta P} ln|P^* - P| \tag{72}$$

where $C = const$ and $\Delta P = |P - P_{SL}|$, $P_{SL} < 0$ is for absolute stability limit hidden in the negative pressures domain.

Notably, 'traditional' (but erroneous) dependence is qualitatively different [129]:

$$m_T(P) \sim V^\#(P) \sim \frac{1}{(P_g^* - P)^\Psi} \tag{71}$$

## 7. The violation of Super-Barus previtreous behavior

The above discussion addressed Super-Arrhenius-type previtreous behavior on cooling and Super-Barus-type behavior on compressing. They are associated with an extreme and systematic previtreous rise of viscosity or slowing-down for relaxation time. However, experimental evolution of $\eta(P)$ or $\tau(P)$ may also exhibit a set of 'anomalous' patterns, namely:

(i) Up to moderate pressures changes $\eta(P)$ or $\tau(P)$ may be weaker than Super-Barus or even basic Barus behavior are observed. On further compressing, returns to SB pattern occurs. It is called the 'inflection' phenomenon. Such a behavior is often observed for a elastohydrodynamic lubrication (EHL), important for machinery applications. [126-123].

(ii) $\eta(P)$ or $\tau(P)$ may decrease during compressing. It can be preceded by almost constant changes or even a slight increase at the low-moderate pressures. Such a behavior is often observed in geophysical relevant systems [130, 131].

(iii) Evolution of $\eta(P)$ or $\tau(P)$ can change from 'fragile' SB behavior to 'strong' one or almost Barus pattern when increasing a temperature of tested isotherm. One can recall glycerol as an example [118, 119, 132-134 ].

Finally, one rises a question, why $\tau(T)$ or $\eta(T)$ changes are described solely by 'strong' or 'fragile' SB behavior?

In this section, the discussion is presented in terms of viscosity changes because phenomena are most often evidenced just for such a physical property. As indicated above, viscosity and primary relaxation time behavior are directly related via Debye-Stokes-Einstein coupling relation [32]. The technological importance of the case (i) leads to heuristic relations that could portray such a behavior. The basic one was proposed in 1952 by McEwan, 1952 [135]:

$$\eta(P) = \eta_0 \exp\left(1 + \frac{P}{(q/a')}\right)^q \tag{73}$$

where $\eta_0, a', q$ are constant parameters.

A few decades later, analysis which recalls Tait classical equation of state for pressure-related changes of volume/density lead to dependence [125]:

$$\eta(P) = A \exp\left[B\ln\left(\frac{C+P}{C+P_r}\right)\right] \tag{74}$$

where $P_r$ denotes the reference pressure. McEwan Eq. (73) may be retrieved from the above relation for $P_r \approx 0$.

Authors of this paper suggest, that McEwan relation may also be derived by the use of extended Avramov-Milchev Eq. (19) relation:

$$\eta(T,P) = \tau_\infty \exp\left(\frac{A_M}{T}\right)^D = \eta_\infty \exp\left(\mu \ln 10 \left(\frac{T_g(P)}{T}\right)^D\right) \tag{75}$$

where $\mu = \log_{10} \tau(T_g, P_g) - \log_{10} \tau_\infty$ is the minimal (reference) fragility. The parallel of this relation can be written for viscosity.

Substituting the Andersson-Andersson (AA) relation [136] for pressure evolution of glass temperature $T_g(P) = T_0(1 + P/a)^{1/b}$ one obtains a relation in agreement with Eq. (74):

$$\eta(P) = \eta_\infty \exp\left(\frac{\mu \ln 10}{T^D}\left(1 + \frac{P}{a}\right)^{D/b}\right) \propto C\left(1 + \frac{P}{a}\right)^{D/b} \tag{77}$$

Recalling analysis of AA relation, which is parallel of Simon-Glatzel dependence used for pressure evolution of melting temperature, the exponent *b* is related to the first derivative of bulk modulus and *a* to bulk modulus itself. Following Eq. (77) such a link may be considered for McEwan Eq. (76), plus the impact of fragility (coefficient $D$ in Eq. 77). For describing $\eta(P)$ or $\eta(P)$ in a broad range of pressures, including the inflection phenomenon, i.e. crossover from McEwan to SB dynamics pattern, Paluch et al. [137, 139] proposed heuristic dependence linking McEwan Eq. 76 and PVFT Eq. (51):

$$\tau(P) = \tau_\infty \exp\left(1 + \frac{P}{(q/a')}\right)^q \exp\left(\frac{D_P P}{P_0 - P}\right) \qquad (78)$$

It was applied successfully for portraying pressure changes of relaxation time, viscosity or electric conductivity upon compressing. Bair proposed to supplement it by the Casalini-Roland (C-R) pressure counterpart by Stickel analysis, $\varphi_P(P) = [dlog_{10}\eta(P)/dP]^{-1/2}$ or $\varphi_P(P) = [dlog_{10}\tau(P)/dP]^{-1/2}$ to validate PVFT behavior at higher pressures. However, such an analysis assumes a priori universality of VFT and PVFT previtreous behavior, which seems to be questionable, as discussed above. A significant problem of Eq. (78) constitute 5 adjustable parameters, leading to a considerable error of parameters in the non-linear fitting.

One can propose alternative 'hybrid' relations, which can be supported by convenient preliminary derivative-based analysis, reducing the number of adjustable parameters to three:

$$\eta(P) = \eta_0 \left(1 + \frac{P}{a_l}\right)^{q_l} \left(1 + \frac{P}{a_h}\right)^{q_h} \qquad (79)$$

where indices '$l$' and '$h$' stand for the low and high pressure domains (below and above the inflection), power exponents $q_l > 0$ and $q_h < 0$ .

Alternatively, one can consider the 'double-critical-like' hybrid relation:

$$\eta(P) = \eta_0 |P_l^* - P|^{\varphi_l} |P_h^* - P|^{\varphi_h} \qquad (80)$$

where power exponents $\varphi_l > 0$ and $\varphi_h < 0$ and $P^*$ denoted the extrapolated singular pressure.

Let's consider linearized derivative-based and distortions-sensitive analysis for leading terms for the above dependencies. For the leading term in critical-type Eq. (80):

$$log_{10}\eta(P) = log_{10}\eta_0 + \varphi log_{10}|P^* - P| \Rightarrow \frac{dlog_{10}\eta(P)}{dP} = \frac{\varphi}{ln10}\frac{1}{|P-P^*|} \Rightarrow \left[\frac{dlog_{10}\eta(P)}{dP}\right]^{-1} = \frac{ln10}{\varphi}|P - P^*| = AP - B \tag{81}$$

and then $\varphi = ln10/A \approx 2.3/A$  and  $P^* = B/A$

For the leading term in McEwan-type Eq. (79):

$$log_{10}\eta(P) = log_{10}\eta_0 + qlog_{10}\left(1+\frac{P}{a}\right) \Rightarrow \frac{dlog_{10}\eta(P)}{dP} = \frac{q}{ln10}\frac{1}{(a+P)/a} \Rightarrow \left[\frac{dlog_{10}\eta(P)}{dP}\right]^{-1} = \frac{ln10}{qa}(a+P) = AP + B \tag{82}$$

and then $q = ln10/A \approx 2.3/A$  and  $a = B/A$

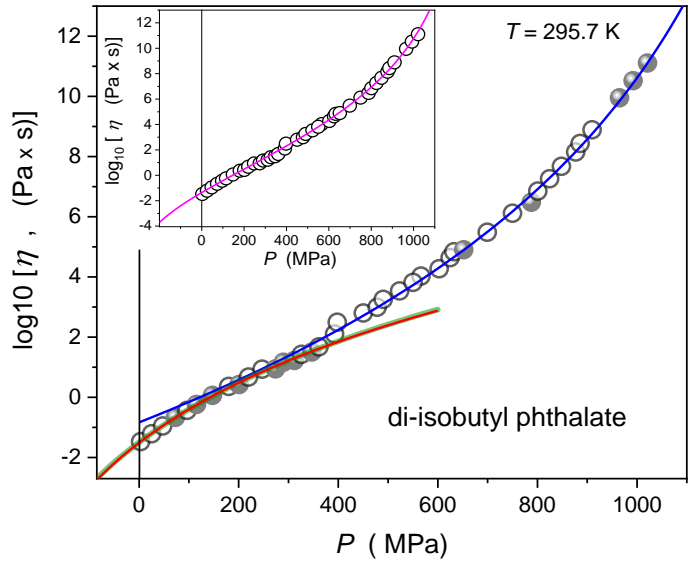

*Fig. 15. Pressure dependence of viscosity in di-isobutyl phtalate (DIIB) for the isotherm $T = 320\ K$. Open circles are related to viscosity measurement [139] supplemented by scaled primary dielectric relaxation time. Full circles are for authors scaled primary relaxation time measurement, supplementing mentioned results. Blue, green and red curve are related to Eqs. (81 and 82) with parameters derived by the derivative-analysis in Fig. 16.  The inset show the parameterization of the double-critical portrayal via Eq. (79), with singular temperatures.*

Hence, the same plot $[dlog_{10}\eta(P)/dP]^{-1}$ vs. $P$ can verify a double-critical-type and double-McEwan-type relations by emergence a linear domain. Linear regression fit can yield optimal values of parameters: $\varphi = q = ln10/A \approx 2.3/A$ and $a = P^* = B/A$. Particularly, the latter is worth stressing, since it is related to $[dlog_{10}\eta(P)/dP]^{-1} = 0$, easily determined graphically. Results of the analysis based on Eqs. (81) and (82) is shown in Figure 15, for compressed di-isobutyl phtalate. The plot shows that the described analysis enable the precise determining of the inflection pressure for the given tested isotherm. Values of related parameters determined via the linear regression fit are also given in the figure. Results presented in Figure 16 have been obtained using experimental data presented in Figure 15 . The main part of the plot shows the fair portrayal of $\eta(P)$ experimental data by single critical-type (Eq. (81)) and McEwan-type (Eq.(82)) terms, separately for both domains, below and above $P_{inf.}$, with parameters given in Fig. 16 to reach domain. The visible overlapping of both type of portrayals for $P < P_{inf.}$ shows that basic McEwan relation is equivalent/isomorhic to critical-like one, with the exponent $\varphi > 0$ and singular pressure $P^* < 0$. For both cases extension into the negative pressure domain, down $P = P^* = a$ is possible.

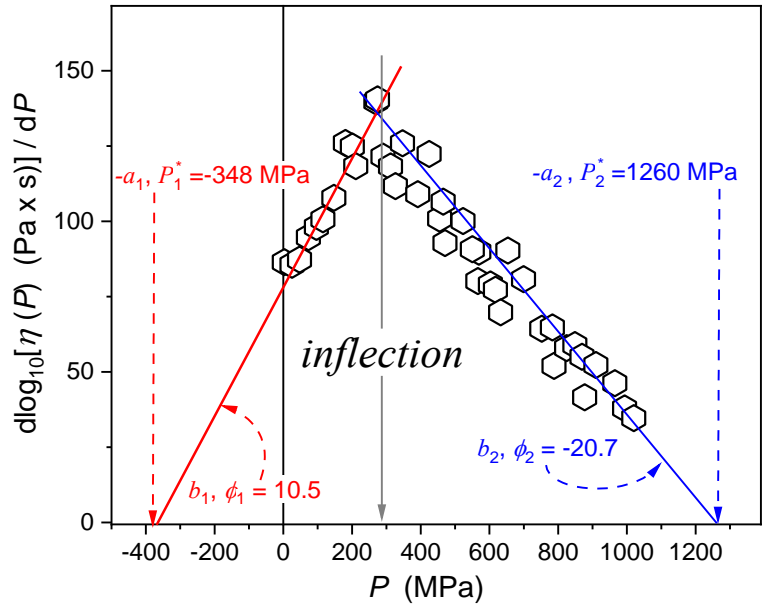

*Fig. 16. Inflection showed based on results of derivative-based analysis related to Eqs. (79) and (80) obtained in di-idobuthyl phtalate taken from Fig. 15.*

Common description of both domains (for pressures below and above the inflection) by 'double; Eq. (79) or Eq. (80) yields the fair portrayal of experimental data, as shown in the inset in Fig. 15. However, a comparable fitting quality can be obtained for parameters different even by 50% or more, particularly regarding the power exponent $q$ or $\varphi$. Results presented in the inset in Fig. 15 are for singular pressure values ($P^*$, $a$). The same ambiguity for parameters appears for the hybrid Eq. (78), proposed by Paluch et. al. [137]. In the opinion of the authors, this essential fitting uncertainty is not related solely to the number of adjustable parameters (5) in Eqs. (78), (79), and (80). It is their inherent feature. As indicated below, infection phenomenon may be considered as a propagation of an impact of a maximum of $T_g(P)$ curve, for tested isotherms located above the maximum. Generally, one may expect that vitrification, at least in liquids, is associated with 'mechanism I' and 'mechanism II'. The latter emerges at extreme pressures and it is associated with the domination of repulsive, hard-sphere-type, interactions. For lower pressures an interplay between attractive and repulsive interaction may lead even to a crossover on compressing. This may give emergence of $T_g(P)$ maximum and following portrayal [132, 133]:

$$T_g(P) = F(P)D(P) = T_g^0 \left(1 + \frac{\Delta P}{\Pi}\right)^{1/b} \exp\left(\frac{P}{c}\right) = T_g^0 \left(1 + \frac{P - P_g^0}{\pi + P_g^0}\right)^{1/b} \exp\left(\frac{P}{c}\right) \tag{83}$$

where $F(P)$ and $D(P)$ are for the rising (SG-type) and damping terms, $\pi < 0$ is for terminal of absolute stability limit pressure at $T = 0$, and $c$ is for damping pressure coefficient. Above relation valid for an arbitrary pressure along $T_g(P)$ curve and may penetrate negative pressures domain. It can even describe systems for which the maximum is hidden in negative pressures domain. It parallel obeys for pressure dependence of melting temperature $T_m(P)$. Worth recalling is a link between both magnitudes, known as the Turnbull criterion: $T_m/T_g \approx w < 1$. For systems particularly 'easily' passing $T_m$ and entering the supercooled, pre-vitreous domain $w \approx 2/3$ is suggested.

Assuming $T_g^0 = T_g(P = 0.1MPa)$ and $P_g^0 = 0.1MPa$ one may approximate the above relation in form first derived by Rein and Demus [137 and therein] and recalled by Kechin:

$$T_g(P) \approx T_g^0 \left(1 + \frac{P}{\pi}\right)^{1/b} \exp\left(\frac{P}{c}\right) \tag{84}$$

It obeys for $P \geq 0$. When neglecting a damping term ($c \to \infty$) it has a form of the Andersson-Andersson (AA) equation for glass temperatures or the Simon-Glatzel relation for melting temperature. For such an approximation $T_g(P)$ and $T_m(P)$ permanently increase on compressing. Notwithstanding, AA- or SG-related approximations may be use below hypothetical maximum of $T_g(P)$ or $T_m(P)$, for systems where $dT_{g,m}(P)/dP > 0$.

In a hybrid Eq. (78) and 'doubled' Eqs. (79) and (80) a low-pressure behavior for $P < P_{inf.}$ is associated with basic McEwan Eq. (73). Following derivation of Eq. (75), its pressure characterization is determined by $T_g(P)$ behavior, i.e., assuming the AA relation as a background and omitting decreases of $T_g(P)$ passing the maximum ($dT_g/dP < 0$). The latter means, that an influence of vitrification 'mechanism I' diminish for $P > P_{inf.}$. However, for basic MacEwan equation impact of 'mechanism I' continuously increases when passing $P_{inf}$, leading to the parasitic bias of fitting results. Consequently, new 'doubled' Eqs. (79, 80) as well as the popular 'hybrid one (Eq. (78)) may be considered only as an effective, practical tool for portrayal $\eta(P)$, $\tau(P)$ or $\sigma(P)$ dependences exhibiting the inflection phenomenon.

Consequently, fundamentally justified seems to be the separate treatment of domain $P < P_{inf.}$ y as single McEwan-type dependence and for $P > P_{inf.}$ by PVFT or critical-like relations with the exponent $\varphi_h < 0$. Leading parameters can be supported by the derivative-based estimations shown in Fig. 16.

However, it is possible to propose a new 'hybrid' relation described the whole range of pressures and coupled to reference values of parameters given by validation preliminary derivative-based analysis (Fig. 15). It can be 'designed' linking Eqs. (77), (80) and (84).

It allows to propose the McEwan-type equation which impact diminishes when passing $P_{inf.}$

$$\eta(P) = \eta_\infty \exp\left(\frac{\mu \ln 10}{T^D}\left(1 + \frac{P}{a}\right)^{D/b} \exp\left(\frac{P}{c/D}\right)\right) \propto C\left(1 + \frac{P}{a}\right)^{D/b} \exp\left(\frac{P}{c/D}\right) \Rightarrow \eta(P) = \eta_0 \left(1 + \frac{P}{a}\right)^{b'} \exp\left(\frac{P}{c'}\right) \quad (85)$$

where $P_{inf.} > P \geq 0$, extension into negative pressures domain requires substitution $P \rightarrow \Delta P = P - P_{SL}$. The latter can be estimated as $P_{SL} \sim a = \pi$

For describing the whole range of pressures, above and below $P_{inf.}$ the following hybrid relation can be considered:

$$\eta(P) = \eta_0 \left(1 + \frac{P}{a}\right)^{b'} \exp\left(\frac{P}{c'}\right) (P_{II}^* - P)^{\varphi_h} \tag{86}$$

for $P > 0$.

Taking into account above considerations an extension into following extension may cover also negative pressures domain:

$$\eta(P) = \eta_0 \left(1 + \frac{P+a}{a}\right)^{b'} \exp\left(\frac{P+a}{c'}\right) (P_{II}^* - P)^{\varphi_h} \tag{87}$$

Eqs. (86) and (87) contains 5 adjustable parameters. For comparison, the hybrid Eq. (78) by Paluch et al. included also 5 ones. However, in Eqs. (86) and (87) four parameters $(a, b')$ and $(P_{II}^*, \varphi_h)$ can be determined from the preliminary derivative-based analysis shown in Fig. 16. The pre-exponential factor $\eta_0 = \eta(P = 0) \approx \eta(P = 0.1 MPa)$, i.e., it may be determined directly from experiment. Consequently, for the final fitting only the coefficient $c'$ (1 parameter) remains. Results of such portrayal based on Eq. (87) are also shown in Fig. 16.

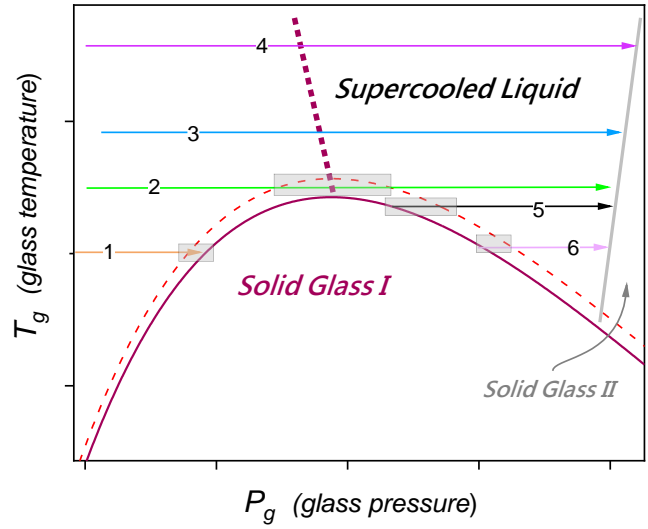

*Fig. 17. Schematic plot of pressure evolution of glass temperature. The glass temperature is represented by $\tau(P_g) = 100s$ and $\eta(P_g) = 10^{13} Poise$ and the brown curve. Red dashed one indicates isochronal/isoviscous curve slightly above glass transition $\sim \tau(P) = 10s$ and $\sim \eta(P) = 10^{12} Poise$. Horizontal colored lines (1-4) indicate paths along which viscosity or relaxation time changes may be tested. Dotted vertical line indicates the suggested 'inflection line' above a maximum of $T_g(P)$ curve. Squares in grey indicates different impacts of pressure changes near glass transition for selected paths.*

Figure 17. schematically shows pressure evolution of glass temperature for a 'model-liquid'. It contains both vitrification mechanisms (I, II) discussed above. The plot indicates basic paths used for $\eta(P)$ or $\tau(P)$ studies. Path (1) is for 'basic' Super-Barus dynamics. For path (3) one expect the appearance of the inflection phenomenon when passing dotted line related to $P_{inf.}$. This is a specific manifestation of a propagation of a maximum of $T_g(P)$ curve impact in over-glass domain of supercooled glass-forming liquid. The 'inflection area' is terminated by paths (4) and (2). For the latter near the top of $T_g(P)$ curve can cause large or even negligible changes of dynamics. When passing the maximum, for $P > P_{inf.}$ one can expect even a slight decrease of $\eta(P)$ or $\tau(P)$ changes until the rise on further compressing, caused by approaching the 'mechanism II' glass transition line. The possibility of the emergence of such 'S-shape' behavior was recently indicated for compressed acetone.

For path 4, the impact of the 'inflection line' diminishes, and one can expect the strong dynamics, close to basic Barus behavior, until approaching the 'mechanism II" vitrification domain, at extreme pressure. The gradual transformation from the 'fragile' to 'strong' $\eta(P)$ or $\tau(P)$ evolution when increasing temperatures of tested isotherms can be found for glycerol, for instance. For paths (5) and (6) one can expect that $\eta(P)$ or $\tau(P)$ first increases on compressing, until the rise associated with approaching the line relate to the second vitrification mechanism appears. For path (6) one can expect that at first the 'flat' domain associated with the proximity od $T_g(P)$ curve maxim before the drop of $\eta(P)$ associated with moving away from 'mechanism I' glass transition curve start to dominate. For some systems the maximum of the $T_g(P)$ can be hidden under negative pressures, and for such systems the behavior associated with $dT_g/dP < 0$ domain can be expected for pressures $P > 0$ (paths 5 and 6). It can explain 'anomalous' patterns of viscosity changes observed in geophysical significant, strongly bonded, magmatic fluids [137].

Note, that none of the above 'anomalous' patterns for $\eta(P)$ or $\tau(P)$ evolutions cannot occur for previtreous $\eta(T)$ or $\tau(T)$ changes, as explicit results from Fig. 16. An exception can be expected for cooling under extreme pressures, close to slowly approaching 'mechanism II" vitrification curve. However, no such experimental results are available yet.

## 8. Conclusions

This report presents a comprehensive discussion of the temperature and pressure-related evolution of dynamic properties such as the primary relaxation time or viscosity in previtreous domain of glass-forming systems. It focuses on taking into account significant distance between available experimentally domain and singular temperature or pressure hidden in solid glass state. It is possible by the use of distortion-sensitive and derivative-based analysis, consequently developed and presented for each model-related scaling equation discussed in this report. Notably, such an analysis leads to decisive conclusions regarding the validity of some important glass transition models.

It is worth stressing, that an extensive discussion of previtreous behavior for pressure path, which also develops new relation and proposes a coherent picture covering classic (Super-Barus) and anomalous (McEwan-type) evolution. It also covers the unique case when relaxation time or viscosity decreases on compressing.

The behavior discussed in a given report may be extended for such properties as diffusion or electric conductivity, but it requires taking into account value of exponent describing the translational-orientational decoupling, as discussed in ref. [32] for instance. This report also indicates limitations in applying popular VFT or WLF relation, which questions the general validity of the Stickel [80] and Casalini-Roland [89] analysis for testing the dynamic crossover phenomenon. This report suggests that reliable tests should be based on the MYEGA equation or the direct analysis via the apparent fragility or the apparent activation energy.

## Acknowledgments

Studies were carried out due to the National Centre for Science (NCN OPUS grant, Poland), ref. UMO-2017/25/B/ST3/02458, headed by Sylwester J. Rzoska, and ref. 2016/21/B/ST3/02203, headed by Aleksandra Drozd-Rzoska. Szymon Starzonek's work was supported by NCN Grant no. 2019/32/T/ST3/00621.

## Authors Information

Aleksandra Drozd-Rzoska, e-mail: arzoska@unipress.waw.pl, ORCID: 0000-0001-8510 2388

Sylwester J. Rzoska, e-mail: sylwester.rzoska@gmail.com, ORCID: 0000-0002-2736-2891

Szymon Starzonek, e-mail: starzoneks@unipress.waw.pl, ORCID: 0000-0003-2793-7971

## Authors Contributions

SJR and ADR are responsible for key conceptual features. SJR, ADR and SS had the basic impact on the preparation of the manuscript. SS and SJR prepared the basic figures.

## Conflict of Interests

The authors declare no conflict of interests.

**References**

1. A. Heyland, L.L. Moroz **Signaling mechanisms underlying metamorphic transitions in animals** *Integr Comp Biol,* 46 (2006), pp. 743-759

2. C.J. Klok **Biological glass: a strategy to survive desiccation and heat** J Exp Biol, 213 (2010), pp. 4-10

3. M. Roskosz, J. Gillot, F. Capet, P. Roussel and H. Leroux, **A sharp change in the mineralogy of annealed protoplanetary dust at the glass transition temperature** A&A, 529 (2011), pp. A111

4. F.B. Wadsworth, A.J. Heap, D.E. Damby, K.-U. Hess, J. Najorka, J. Vasseur, D. Fahner, D.B. Dingwell **Local geology controlled the feasibility of vitrifying Iron Age buildings** Sci Rep, 7 (2017), pp. 40028

5. J. Pagacz, P. Stach, L. Natkaniec-Nowak, B. Naglik, A.P. Drzewicz **Preliminary thermal characterization of natural resins from different botanical sources and geological environments** J Therm Anal Calorim 138 (2019), pp. 4279-4288

6. M. Petters, S. Kasparoglu **Predicting the influence of particle size on the glass transition temperature and viscosity of secondary organic material** Sci Rep 10 (2020), pp. 15170

7. Y.H. Roos **Glass transition temperature and its relevance in food processing** Ann Rev Food Sci Technol 1 (2010), pp. 469-495

8. N.R. Yadhav, V.L. Gaikwad, K. . Nair, H.M. Kadam **Glass transition temperature: Basics and application in pharmaceutical sector** Asian J Pharm 2 (2009), pp. 82-89

9. S. Sahu, S. Saraf **Bioengineering techniques for the efficacy of herbal cosmetics** Res. J. Topical and Cosmetic Sci 1 (2010), pp.1-12

10. H.T. Hoang Nguyen, P. Qi, M. Rostagno, A. Feteha, S.A. Miller **The quest for high glass transition temperature bioplastics J Mater Chem A** 6 (2018), pp. 9298-9331

11. K. Januchta, R. E. Youngman, A. Goel, M. Bauchy, S. L. Logunov, S. J. Rzoska, M. Bockowski, L. R. Jensen, M. M. Smedskjaer **Discovery of ultra-crack-resistant oxide glasses with adaptive networks** ACS Chem Mat 29 (2017), pp. 5865-5876

12. P. Baranowski, S. Starzonek, A. Drozd-Rzoska, S.J. Rzoska, M. Boćkowski, P. Keblinski, T.K. Pietrzak, J.E. Garbarczyk **Multifold pressure-induced increase of electric conductivity in LiFe$_{0.75}$V$_{0.10}$PO$_4$ glass** Sci Rep 9 (2019), pp. 16607

13. D.E. Pegg, **Principles of cryopreservation** Methods Mol Biol 368 (2007), pp. 39-57

14. M.A. Anisimov **Critical Phenomena in Liquids and Liquid Crystals** Gordon and Breach, Reading: 1992.

15. J. Honig, J. Spałek **A Primer to the Theory of Critical Phenomena** Elsevier, Amsterdam: 2018

16. J. Chrapeć, S.J. Rzoska, and J. Zioło **Pseudospinodal curve for binary solutions determined from the nonlinear dielectric effect** Chem Phys 111 (1987), pp. 155-160

17. S. J. Rzoska, A. Drozd-Rzoska, P.K. Mukherjee, D.O. Lopez, J. C. Martinez Gracia, **Distortion-sensitive insight into the pretransitional behavior of 4-n-octyloxy-4′-cyanobiphenyl (8OCB)** J Phys: Condens Matt 25 (2013), pp. 245105

18. A Drozd-Rzoska, S Rzoska, S Pawlus, J Zioło **Complex dynamics of supercooling n-butylcyanobiphenyl (4CB)**, Phys Rev. E 72 (2005), pp. 031501

19. A Drozd-Rzoska, SJ Rzoska, J Zioło **Mean-field behaviour of the low frequency non-linear dielectric effect in the isotropic phase of nematic and smectic n-alkylcyanobiphenyls** Liquid Crystals 21 (1996) pp. 273-277

20. A. Drozd-Rzoska, S.J. Rzoska, and K. Czupryński **Phase transitions from the isotropic liquid to liquid crystalline mesophases studied by "linear" and "nonlinear" static dielectric permittivity** Phys. Rev. E 61 (2000), pp. 5355-5360

21. A. Drozd-Rzoska, S.J. Rzoska, A. Szpakiewicz-Szatan, J. Łoś, K. Orzechowski *Pretransitional and premelting effects in menthol* *Chem Phys Lett* 793 (2022), pp. 139461

22. A. Drozd-Rzoska, S.J. Rzoska, A. Szpakiewicz-Szatan, S. Starzonek, J. Łoś, K. Orzechowski *Supercritical anomalies in liquid ODIC-forming cyclooctanol under the strong electric field* *J. Mol. Liq.* 345 (2022), pp. 1178491

23. D. Kennedy, C. Norman **What don't we know. Science's 125 Open Questions; in 125[th] anniversary** Science 1[st] July special issue (2005)

24. L. Berthier, M. Ediger **Facets of the glass transition** Physics Today 69 (2016), pp. 40-44

25. F. Kremer, A. Loidl **Scaling of Relaxation Processes** Springer, Berlin (2018)

26. R. Ramirez **An Introduction to Glass Transition** Nova Sci Pub, London: (2019)

27. G.B. McKenna Glass transition: challenges of extreme time scales and other interesting **problems** Rubber Chem and Technol 93 (2020) pp. 79-120

28. C.A. Angell **Formation of glasses from liquids and biopolymers** Science 267 (1995) pp. 1924-1935

29. R. Böhmer, K.L. Ngai, C.A. Angell, D.J. Plazek **Nonexponential relaxations in strong and fragile glass formers** J Chem Phys 99 (1993) pp. 4201-4209

30. J.C. Mauro and R.J. Loucks **Impact of fragility on enthalpy relaxation in glass** Phys. Rev. E 78 (2008), pp. 021502

31. L.-M. Wang, J. Mauro **An upper limit to kinetic fragility in glass-forming liquids** J Chem Phys 134 (2011), pp. 044522

32. S. Starzonek, S.J. Rzoska, A. Drozd-Rzoska, S. Pawlus, J.-C. Martinez-Garcia, L. Kistersky **Fractional Debye–Stokes–Einstein behaviour in an ultraviscous nanocolloid: glycerol and silver nanoparticles** Soft Matter 5 (2011) pp. 5554-5562

33. J.C. Martinez Garcia, J.Ll. Tamarit, S.J. Rzoska **Enthalpy space analysis of the evolution of the primary relaxation time in ultraslowing systems** J Chem Phys 134 (2011), pp. 024512

34. M.L. Williams, R.F. Landel, and J.D. Ferry **The temperature dependence of relaxation mechanisms in amorphous polymers and other glass-forming liquids** J Am Chem Soc 77 (1955), pp. 3701-3706

35. D. Turnbull, M.H. Cohen **Free-volume model of the amorphous phase: glass transition** J Chem Phys 34 (1961), pp. 120-124

36. G. Adam, and J.H. Gibbs **On the temperature dependence of cooperative relaxation properties in glass-forming liquids** J Chem Phys 43 (1965), pp. 139-146

37. H. Tanaka, T. Kawasaki, H. Shintani, K. Watanabe **Critical-like behaviour of glassforming liquids** Nat Mater 9 (2010), pp. 324-331

38. L. Wang. C.A. Angell, R. Richert **Fragility and thermodynamics in nonpolymeric glass-forming liquids** J Chem Phys 125 (2006), pp. 074505

39. O.N. Senkov, D.B. Miracle **Correlation between thermodynamic and kinetic fragilities in nonpolymeric glass-forming liquids** J Chem Phys 128 (2008),124508

40. H. Tanaka **Relation between thermodynamics and kinetics of glass-forming liquids** Phys Rev Lett 90 (2003), pp. 05570

41. C.-S. Lee, M. Lulli, L.-H. Zhang, H.-Y. Deng, C.-H. Lam **Fragile glasses associated with a dramatic drop of entropy under supercooling** Phys Rev Lett 125 (2020), pp. 265703

42. R. Richert, W. H. Woodward, T. Fielitz, C. Todd **Using derivative plots to ascertain fragilities of glass-formers** J Non -Cryst Solids 553 (2021), pp. 120478

43. J.C. Dyre, N.B. Olsen *Landscape equivalent of the shoving model* Phys Rev E 69 (2004), pp. 042501

44. T. Hecksher, A.I. Nielsen, N.B. Olsen, and J.C. Dyre **Little evidence for dynamic divergences in ultraviscous molecular liquids** Nat Phys 4 (2008), pp. 737-741

45. J. C. Martinez Garcia, S. J. Rzoska, A. Drozd-Rzoska, J. Martinez Garcia **A universal description of ultraslow glass dynamics** Nat Comm 4 (2013), 4, pp. 1823

46. J. C. Martinez Garcia, S.J. Rzoska, A. Drozd-Rzoska, J. Martinez Garcia, and J.C. Mauro, **Divergent dynamics and the Kauzmann temperature in glass forming systems** Sci Rep 4 (2014), pp. 5160

47. J.C. Martinez Garcia, S.J. Rzoska, A. Drozd-Rzoska, S. Starzonek, and J.C. Mauro **Fragility and basic process energies in vitrifying system** Sci Rep 5 (2015), pp. 8314

48. G.B. McKenna **Diverging views on glass transition** Nat Phys 4 (2008), pp. 673

49. K.L. Ngai **Relaxation and Diffusion in Complex Systems** Springer, Berlin: 2010

50. M.I. Ojovan **On viscous flow in glass-forming organic liquids** Molecules 25 (2020), pp. 4029

51. J.-H. Hung, T.K. Patra, D.S. Simmons **Forecasting the experimental glass transition from short time relaxation data** J Non-Cryst Solids 544 (2020), 120205

52. J.C. Mauro, Y. Yue, A.J. Ellison, P.K. Gupta, D.C. Allan **Viscosity of glassforming liquids** Proc Natl Acad Sci USA 24 (2009), pp. 19780–19784

53. I. Avramov, A. Milchev **Effect of disorder on diffusion and viscosity in condensed systems** J Non-Cryst Solids 104 (1988), pp. 253-260

54. L.S. Garcia-Colin, L.F. del Castillo, and P. Goldstein **Theoretical basis for the Vogel-Fulcher-Tammann equation** Phys Rev B 40 (1990), pp. 7040-7444

55. R. Richert **Scaling vs. Vogel-Fulcher-type structural relaxation in deeply supercooled materials** Physica A 287 (2000), pp. 26-36

56. R.H. Colby **Dynamic scaling approach to glass formation** Phys Rev E 61 (2000), pp. 1783-1792

57. B.M. Erwin, R.H. Colby, **Temperature dependence of relaxation times and the length scale of cooperative motion for glass-forming liquids** J Non-Cryst Solids 307-310 (2002), pp. 225-232

58. M. Paluch. Z. Dendzik and S.J. Rzoska **Scaling of high-pressure viscosity data in low-molecular-weight glass-forming liquids** Phys Rev B 60 (1999), pp. 2979-2982

59. Y.S. Elmatad, D. Chandler, J.P. Garrahan **Corresponding states of structural glassformers** J Phys Chem B 113 (2009), pp. 5563-5567

60. M.L Ferreira Nascimento, C. Aparicio **Data classification with the Vogel–Fulcher–Tammann–Hesse viscosity equation using correspondence analysis** Physica B 398 (2007), pp. 71-77

61. F. Mallamace, C. Branca, C. Corsaro, N. Leone, J. Spooren, S.-H., Chen, H.E. Stanley, **Transport properties of glass-forming liquids suggest that dynamic crossover temperature is as important as the glass transition temperature** Proc Natl Acad Sci USA 107 (2010), pp. 22457-22462

62. N. Weingartner, C. Pueblo, F. S. Nogueira, K.F. Kelton, Z. Nussinov, **A phase space approach to supercooled liquids and a universal collapse of their viscosity** Front Mat 3 (2016), pp. 50

63. M. E. Blodgett, T. Egami, Z. Nussinov, K. F. Kelton **Proposal for universality in the viscosity of metallic liquids** Sci Rep 5 (2015) pp. 13837

64. S.A. Arrhenius **Über die Dissociationswärme und den Einfluß der Temperatur auf den Dissociationsgrad der Elektrolyte** Z Phys Chem 4 (1889), pp. 96–116

65. J. de Guzman **Relación entre la Fluidez y el Calor de Fusion** Anales de la Sociedad Espanola de Fisica y Quimica 11 (1913), pp. 353-362

66. C.V. Raman **A theory of the viscosity of liquids** Nature 111 (1923), pp. 532-533

67. C. Andrade **A Theory of the Viscosity of Liquids. - Part I.** London Edinb Dub Philos Mag J Sci 17 (1934), pp. 497–511

68. G. Williams **Complex dielectric constant of dipolar compounds as a function of temperature, pressure and frequency** Trans Faraday Soc 60 (1964), pp. 1548-1555

69. G. Williams **Complex dielectric constant of dipolar compounds as a function of temperature, pressure and frequency. Part 2.—The α–relaxation of polymethyl acrylate** Trans. Faraday Soc 60 (1964) pp. 1556-1573

70. H. Vogel **Temperaturabhängigkeitsgesetz der viskosität von flüssigkeiten** Phys Zeit 22 (1921) pp. 645-646

71. G.S. Fulcher **Analysis of recent measurements of the viscosity of glasses** J Am Ceram Soc 8 (1925) pp. 339-335

72. G, Tammann **Glasses as supercooled liquids** J Soc Glass Technol 9 (1925) pp. 166-185.

73. A. K. Doolittle **Studies in Newtonian Flow. II. The Dependence of the Viscosity of Liquids on Free-Space** J Appl Phys 22 (1951), pp. 1471–1475

74. T.G. Fox, P.J. Flory **Second-order transition temperatures and related properties of polystyrene", Journal of Applied Physics** 21 (1950), pp. 581–591

75. L. Berthier, M. Ozawa, C. Scaillet **Configurational entropy of glass-forming liquids**, J Chem Phys 150 (2019), pp. 160902

76. J.P. Garrahan, D. Chandler **Coarse-grained microscopic model of glass formers** Proc Natl Acad Sci USA 100 (2003), pp. 9710-9714

77. D. Chandler, J.P. Garrahan **Thermodynamics of coarse-grained models of supercooled liquids** J Chem Phys 12 (2005), pp. 044511

78. M.M. Smedskjaer, J.C. Mauro, Y.Z. Yue **Ionic diffusion and the topological origin of fragility in silicate glasses** J Chem Phys 131 (2009), pp. 244514

79. J. T. Bendler and M. F. Shlesinger **Generalized Vogel law for glass-forming liquids** J Stat Phys 53 (1988), pp. 531-541

80. F. Stickel, E.W. Fisher, and R. Richert **Dynamics of glass-forming liquids. I. Temperature-derivative analysis of dielectric relaxation data** J Chem Phys 102 (1995), pp. 6251-6257

81. S. Corezzi, M. Beiner, H. Huth, K. Schröter, S. Capaccioli, R. Casalini, E. Donth **Two crossover regions in the dynamics of glass forming epoxy resins** J Chem Phys 117 (2002), pp. 2435-2446

82. W. Götze **The essentials of the mode-coupling theory for glassy dynamics** Cond Mat Phys 1 (1998), pp. 873–904

83. W. Götze, R. Schilling **Glass transitions and scaling laws within an alternative mode-coupling theory** Phys Rev E 91 (2015), pp. 042117

84. V.N. Novikov, and A.P. Sokolov **Universality of the dynamic crossover in glassforming liquids: a "magic" relaxation time** Phys Rev E 67 (2003), pp. 031507

85. C.M. Roland **Characteristic relaxation times and their invariance to thermodynamic conditions** Soft Matter 4 (2008), pp. 2316-2322.

86. A. Drozd-Rzoska **A universal behavior of the apparent fragility in ultraslow glass forming systems** Sci Rep 9 (2019) pp. 6816

87. A. Drozd-Rzoska, S.J. Rzoska **On the derivative-based analysis for temperature and pressure evolution of dielectric relaxation times in vitrifying liquids** Phys. Rev. E 73 (2006) pp. 041502.

88. J.C. Dyre **Colloquium: The glass transition and elastic models of glass-forming liquids** Rev Mod Phys 79 (2006), pp. 953-972

89. R. Casalini, and C. M. Roland **Viscosity at the dynamic crossover in o-terphenyl and salol under high pressure** Phys Rev Lett 92 (2004), pp. 245702-245706

90. S.S.N. Murthy **Experimental study of dielectric relaxation in supercooled alcohols and polyols** Molecular Physics 87 (1996), pp. 691-709

91. F. Kremer, A. Schönhals **Broadband Dielectric Spectroscopy**, Springer, Berlin: 2004

92. M. Naoki, K. Satoshi, **Contribution of hydrogen bonds to apparent mobility in supercooled D-sorbitol and some polyols** J Phys Chem 96 (1991), pp. 431-437

93. Y. Suzuki **Effect of OH groups on the polyamorphic transition of polyol aqueous solutions** J Chem Phys 150 (2019), pp. 224508

94. H. Bässler **Viscous flow in supercooled liquids analyzed in terms of transport theory for random media with energetic disorder** Phys Rev Lett 58 (1987), pp. 767-770

95. A. Drozd-Rzoska, S.J. Rzoska, and C.M Roland **On the pressure evolution of dynamic properties of supercooled liquids** J Phys: Condens Matt 20 (2008), pp. 244103.

96. N. Canorini, F. Martinelli, C. Roberto, C. Toninelli **Kinetically constrained models**, pp. 741-752; in V. Sidoravičius (ed.) New Trends in Mathematical Physics, Springer, Berlin: 2009

97. A.S. Keys, L.O. Hedges, J.P. Garrahan, S. C. Glotzer, D. Chandler **Excitations are localized and relaxation is hierarchical in glass-forming liquids** Phys Rev X 1 (2013), pp. 21013

98. C. P. Royal, F. Turci, T. Speck **Dynamical phase transitions and their relation to structural and thermodynamic aspects of glass physics** J Chem Phys 153 (2020), 090901

99. S. Karmakara, C. Dasguptaa, S. Sastry **Growing length and time scales in glass-forming liquids** Proc Natl Acad Sci USA 106 (2009), pp. 3575-3579

100. A. Dehaoui, B. Issenmann, F. Caupin, **Viscosity of deeply supercooled water and its coupling to molecular diffusion** Proc Natl Acad Sci USA 29 (2015), pp. 12020-12025

101. P.C. Hohenberg and B.I. Halperin **Theory of dynamic critical phenomena** Rev Mod Phys. 49 (1977), pp. 435-479

102. J. Souletie, and D. Bertrand **Glasses and spin glasses: a parallel** J Phys (France) 11 (1991), pp. 1627-1637

103. J. Souletie **Hierarchical scaling: An analytical approach to slow relaxations in spin glasses, glasses, and other correlated systems** J Appl Phys 75 (1994), pp. 5512-5516

104. E.J. Saltzman and K.S. Schweizer **Universal scaling, dynamic fragility, segmental relaxation, and vitrification in polymer melts** J Chem Phys 121 (2004), pp. 2001-2009

105. A. Drozd-Rzoska **Heterogeneity-related dynamics in isotropic n-pentylcyano biphenyl** Phys Rev E 73 (2006) pp. 022501

106. A. Drozd-Rzoska, S.J. Rzoska, S. Pawlus, J.C. Martinez-Garcia, and J.-L. Tamarit **Evidence for critical-like behavior in ultraslowing glass-forming systems** Phys Rev E 82 (2010), 031501

107. J.C. Martinez-Garcia, J. Martinez-Garcia and S.J. Rzoska **The new insight into dynamic crossover in glass forming liquids from the apparent enthalpy analysis** J Chem Phys 137 (2012), pp. 064501

108. A. Drozd-Rzoska **Glassy dynamics of liquid crystalline 4'-n-pentyl-4-cyanobiphenyl (5CB) in the isotropic and supercooled nematic phases** J Chem Phys 130 (2009), pp. 234910

109. A. Drozd-Rzoska **'Quasi-Tricritical' and Glassy Dielectric Properties of a Nematic Liquid Crystalline Material**, Crystals 10 (2020), 297

110. S. Starzonek, A. Drozd-Rzoska, S.J Rzoska, K. Zhang, E. Pawlikowska, A. Kędzierska-Sar, M. Szafran, F. Gao **Polivinylidene difluoride-based composite: unique glassy and pretransitional behavior** Europ Phys J B 93 (2019), pp. 55

111. R. Levit, J.C. Martinez-Garcia, D.A. Ochoa, J. E. Garcia **The generalized Vogel-Fulcher-Tamman equation for describing the dynamics of relaxor ferroelectrics** Sci Rep 9 (2019), pp. 12390

112. P.G. Wolyness, J.-P. Bouchaud **Structural Glasses and Supercooled Liquids: Theory, Experiment, and Applications**; Wiley & Sons Ins, NY: 2012

113. A. Drozd-Rzoska, S.J. Rzoska, S. Starzonek **New paradigm for configurational entropy in glass-forming systems** Sci Rep 12, 3058 (2022)

114. C. Barus **Isothermals, isopiestics and isometrics relative to viscosity** Am J Sci 45 (1893) pp. 87-96

115. S.J. Rzoska **New Challenges for the Pressure Evolution of the Glass Temperature** Front Mat: Glass Sci 4 (2017), pp. 33

116. C.M. Roland, S. Hensel-Bielowka, M. Paluch, R. Casalini **Supercooled dynamics of glass-forming liquids and polymers under hydrostatic pressure** Rep Prog Phys 68 (2005), pp. 1405

117. G. Floudas, M. Paluch, A. Grzybowski, K. Ngai **Molecular Dynamics of Glass-Forming Systems: Effects of Pressure**; Springer, Berlin, 2011

118. G.P. Johari, and E.P. Whalley **Dielectric Properties of glycerol in the range 0.1-105 Hz, 218-357 K, 0-53 kbar** Faraday Symp Chem Soc 6 (1972), pp. 23–41

119. M. Paluch, S.J. Rzoska, and J. Zioło **On the pressure behaviour of dielectric relaxation times in supercooled, glassforming liquids** J Phys: Condens Matt 10 (1998), pp. 4131–4138

120. A. R. Imre, H. J. Maris, and P. R. Williams **Liquids under Negative Pressures** Kluwer, Dordrecht:2002

121. A. R. Imre, A. Drozd-Rzoska, A. Horvath, Th. Kraska, S. J. Rzoska **Solid-fluid phase transitions under extreme pressures including negative ones** J Non-Cryst Solids 354 (2008), pp. 4157-4162

122. C.A. Angell, and Z. Quing, **Glass in a stretched state formed by negativepressure vitrification: trapping in and relaxing out** Phys Re. B 39 (1989), pp. 8784–8787

123. C.J.A. Roelands **Correlational aspects of the viscosity-temperature-pressure relationship of lubricating oils** Doctoral dissertation Delft University of Technology, Delft: 1966

124. S. Bair **Roelands' missing data** Proc Inst Mech Engn, Part J: J Engn Tribol 218 (2004), pp. 57-60

125. S. Bair **High Pressure Rheology for Quantitative Elastohydrodynamics** Elsevier Amsterdam: 2019

126. S. Bair, M. Baker, D. M. Pallister **Revisiting the compressibility of oil/refrigerant lubricants** J Tribology 139 (2017), pp. 024501

127. S. Kießkalt **Untersuchungen über den Einfluß des Druckes auf die Zähigkeit von Ölen und seine Bedeutung auf die Schmiertechnik** VDI-Forsch.-Heft 291 Berlin: 1927

128. A. Drozd-Rzoska **Pressure-Related Universal Previtreous Behavior of the Structural Relaxation Time and Apparent Fragility** Front Mat: Glass Sci. 6, (2019). pp. 103

129. A. Drozd-Rzoska **Activation volume in superpressed glass formers** Sci Rep 9 (2019), pp. 13787

130. C.M. Scarfe, B.O. Mysen and D.L. Virgo Pressure dependence of the viscosity of silicate melts. Geocemical Society, Special Publication No. 1 (1987), pp. 59–67

131. G. Schubert **Treates on Geophysics** (ed.) Elsevier, Amsterdam: 2015; D.B. Dingwell **Properties of rocks and minerals – diffusion, viscosity, and flow of melts**, p. 473

132. A. Drozd-Rzoska **Pressure dependence of the glass temperature in supercooled liquids** Phys Rev E 72 (2005), pp. 041505

133. A. Drozd-Rzoska, S. J. Rzoska, and M. Paluch **On the glass transition under extreme pressure** J Chem Phys 126 (2007), pp. 164504

134. A.A. Pronin, M.V. Kondrin, A. G. Lyapin, V.V. Brazkhin, A.A. Volkov, P. Lunkenheimer **Glassy dynamics under superhigh pressure** Phys. Rev. E 81(2010), pp. 041503

135. E. McEwen **The effect of variation of viscosity with pressure on the load-carrying capacity of the oil film between gear-teeth** J Inst Petroleum 38 (1952), pp. 646–672

136. S.P. Andersson, and O. Andersson **Relaxation studies of poly(propylene glycol) under high pressure** Macromolecules 31 (1998), pp. 2999–3006

137. E. Thoms, Z. Wojnarowska, P. Goodrich, J. Jacquemin, and M. Paluch **Inflection in the pressure dependent conductivity of the protic ionic liquid $C_8HIM$ $NTf_2$** J Chem Phys **146** (2017) pp. 181102

138. S. Bair, L. Martinie, P. Vergne Classical EHL versus quantitative EHL: A perspective Part II – Super-Arrhenius piezoviscosity, an essential component of electrohydrodeynamic friction missing from classical EHL **Tribol Lett** 63 (2016) 38

139. M. Sekuła, S. Pawlus, S. Hensel-Bielowka, J. Zioło, M. aluch, C.M. Rolan.Structural and secondary relaxations in supercooled di-n-butyl and diisobutyl phthalate at elevated pressures **J Phys Chem** B 108 (2004) 4997-5003

140. A.G.M. Ferreira, A.P.V. Egas , I.M.A. Fonseca, A.C. Costa, D.C. Abreu, L.Q. Lobo **The viscosity of glycerol** J Chem Thermod 113 (2017) 162-182